\documentclass{article}

\usepackage{amssymb}
\usepackage{amsmath}
\usepackage{graphicx}
\usepackage{epsfig}
\usepackage{float}
\usepackage{subfigure}
\usepackage{psfrag}
\usepackage[T1]{fontenc}
\usepackage{color}
\usepackage{array}
\usepackage{dsfont}
\usepackage{a4wide}
\usepackage{algorithm,algorithmic}
\usepackage{hyperref}
\hypersetup{backref,colorlinks=true,linkcolor=blue,citecolor=blue}
\usepackage{epsfig}
\usepackage{listings}
\definecolor{greenMatlab}{RGB}{46,139,87} % vert Matlab
\usepackage{tikz}
\usetikzlibrary{arrows,decorations.pathmorphing,backgrounds,fit,positioning,shapes.symbols,chains,shapes}

\title{Yet another introduction to linear dynamical systems control:  From identification and approximation to digital control}
\author{C. Poussot-Vassal and P. Vuillemin}
\date{March 2020}

\definecolor{bleuONERA}{RGB}{16,97,169}
\definecolor{grisONERA}{RGB}{64,64,66}
\providecommand{\blue}[1]{\textcolor[RGB]{16,97,169}{#1}} % Bleu charte ONERA
\providecommand{\red}[1]{\textcolor[rgb]{0.98,0.00,0.00}{#1}}

\providecommand{\code}[1]{\textbf{\texttt{#1}}} %
\providecommand{\codeScript}[1]{\blue{\textbf{\texttt{#1}}}} %

\providecommand{\tf}[1]{\textbf{\texttt{TF}}\pare{#1}} %

\providecommand{\mor}[0]{\textbf{MOR}~}

\providecommand{\matlab}[0]{\textsc{Matlab}~}

\providecommand{\mimo}[0]{\textbf{MIMO}~}

\providecommand{\siso}[0]{\textbf{SISO}~}
\providecommand{\dae}[0]{\textbf{DAE}~}
\providecommand{\ode}[0]{\textbf{ODE}~}

\providecommand{\lti}[0]{\textbf{LTI}~}

\providecommand{\ie}[0]{\emph{i.e.}~}

\providecommand{\eg}[0]{\emph{e.g.}~}

\newenvironment{eq}{\everymath {\displaystyle \everymath{ }} \equation}{ \endequation} %
\providecommand{\abs}[1]{\left\lvert #1 \right\rvert} %
\providecommand{\norm}[1]{|| #1 ||} %
\providecommand{\eval}[2]{\left.#1\right\rvert_{#2}} %
\providecommand{\rank}[0]{\mathbf{rank} } %
\providecommand{\pare}[1]{\left(#1\right) } %
\providecommand{\x}[0]{\mathbf{x}} %
\renewcommand{\u}{\mathbf{u}} %
\providecommand{\y}[0]{\mathbf{y}} %
\providecommand{\yr}[0]{\mathbf{\hat{y}}} %
\providecommand{\Htranr}[0]{\mathbf{\hat{H}}} %
\providecommand{\Er}[0]{{\hat{E}}} %
\providecommand{\Ar}[0]{{\hat{A}}} %
\providecommand{\Br}[0]{{\hat{B}}} %
\providecommand{\Cr}[0]{{\hat{C}}} %

\providecommand{\Hsystem}[0]{\mathbf{\Sigma}} %
\providecommand{\Hreal}[0]{\mathcal{S}} %
\providecommand{\Htran}[0]{\mathbf{H}} %
\providecommand{\Gtran}[0]{\mathbf{G}} %
\providecommand{\E}[0]{{E}} %
\providecommand{\A}[0]{{A}} %
\providecommand{\B}[0]{{B}} %
\providecommand{\C}[0]{{C}} %
\providecommand{\D}[0]{{D}} %

\providecommand{\Htwo}[0]{{\mathcal{H}_{2}}} %
\providecommand{\Hinf}[0]{{\mathcal{H}_{\infty}}} %

\providecommand{\Cplx}[0]{\mathds{C}} %
\providecommand{\Real}[0]{\mathds{R}} %
\providecommand{\matrixtwo}[4]{ \left[\begin{array}{cc} #1 & #2 \\ #3 & #4 \end{array}\right] } %
\providecommand{\vectortwo}[2]{ \left[\begin{array}{c} #1 \\ #2 \end{array}\right] } %
\providecommand{\vectorthree}[3]{ \left[\begin{array}{c} #1 \\ #2 \\ #3 \end{array}\right] } %
\providecommand{\vectorfour}[4]{ \left[\begin{array}{c} #1 \\ #2 \\ #3 \\ #4 \end{array}\right] } %
\providecommand{\vectortwoT}[2]{ \left[\begin{array}{cc} #1 & #2 \end{array}\right] } %

\begin{document}

\maketitle

\begin{abstract}
This report aims at presenting (yet) a(nother) methodology to design and implement a linear controller for linear dynamical systems on practical applications. The specificity of this report is that authors try to cover (obviously in a non exhaustive way) a wide range of control engineering fields. Indeed, the main purpose is to give a quick overview  of standard control engineer approaches to non familiar readers. More specifically, using a simple toy example, we discuss the main steps control engineers usually follow. Namely, \emph{(i)} the excitation signals construction, \emph{(ii)} the (continuous-time) linear model construction and approximation, \emph{(iii)} the (continuous-time) control design, and finally, \emph{(iv)} its time-domain discretisation and control signal modulation in view of practical implementation. This report is clearly user-oriented and thus focuses on practical aspects (using \matlab code) rather than on theoretical ones, let to the reader's curiosity with few but relevant references.
\end{abstract}

\section{Motivation and framework}
\label{sec-motivation}
As stated in the abstract, we will consider dynamical systems and control design through the lens of continuous-time linear functions. Readers should keep in mind that nonlinear systems theory and methods exist but, to the authors feeling, may be viewed as more tedious to apply in a systematic way\footnote{Beside, they are far from the authors know-how.}. The discrete-time (or sampled-time) will be briefly discussed in the last section.

\subsection{Assumption and framework}

%stable\footnote{Unstable systems are classically involved in the dynamical and control developments but are more complex to consider in the identification one. This is why stable systems are considered here.}
Generally speaking, this report considers finite order Multiple Input Multiple Outputs (\mimo)  $n_u$ inputs $n_y$ outputs Linear Time Invariant (\lti) dynamical systems denoted $\Hsystem$ described by a complex, or frequency-domain, transfer function $\Htran$,
\begin{equation}
\Htran : \Cplx \rightarrow \Cplx^{n_y \times n_u},
\end{equation}
which maps linearly the inputs $\u$ to the outputs $\y$ as,
%which - complex or frequency-domain - transfer function may be described by a set of first order \lti Ordinary Differential Equations (\ode):
\begin{equation}
\begin{array}{rcl}
\y(s) &=& \Htran(s) \u(s) \\
\end{array}
\label{eq:H}
\end{equation}
This mapping can also be seen from a time-domain perspective as a set of first-order Differential Algebraic Equations (\dae{}) described by a descriptor realisation $\Hreal$,
%for which time-domain descriptor realisation $\Hreal$ reads
\begin{equation}
\E \dot{\mathbf x}(t) =\A \x(t) +\B \u(t)  \text{  and  } \y(t) =\C \x(t) ,
\label{eq:S}
\end{equation}
where $\x(t) \in \Real^{n}$ denotes the internal variables (the state variables if $\E$ is invertible), $\u(t) \in \Real^{n_u}$ and $\y(t) \in \Real^{n_y}$ are the input and output signals, respectively, and $\E,\A \in \mathbb{R}^{n \times n}$, $\B \in \mathbb{R}^{n\times n_u}$ and $\C \in \mathbb{R}^{n_y \times n}$ are constant matrices. It is common to name $(\E,\A)$ as the dynamical matrices, while $\B$ and $\C$ are the input and output ones.

\begin{figure}[h]
    \centering
    \scalebox{1}{\pgfdeclarelayer{background}
\pgfdeclarelayer{foreground}
\pgfsetlayers{background,main,foreground}

% Define a few styles and constants
\tikzstyle{sysProp}  = [draw=black,thick, fill=bleuONERA!20,text width=12em,text centered,minimum height=8em,rounded corners]

\begin{tikzpicture}

%% SYSTEM
\node (sys) [sysProp] {Dynamical system $\Hsystem$ \\ described by $\Htran$ and $\Hreal$\\ $\x=\vectorthree{x_1(.)}{x_2(.)}{\vdots}$};

%% INPUT OUTPUT ARROWS
\path (sys.0)+(2.5cm,0) node (outputM) [] {$\vectorfour{y_1(.)}{y_2(.)}{\vdots}{y_{n_y}(.)}=\y$};
\draw [->,line width=1.5pt] (sys.0) -- (outputM);
\path (sys.0)+(-7cm,0) node (inputW) [] {$\u=\vectorfour{u_1(.)}{u_2(.)}{\vdots}{u_{n_u}(.)}$};
\draw [->,line width=1.5pt] (inputW) -- (sys.180);

\end{tikzpicture}}
    \caption{Graphical view of a (linear) dynamical system with inputs $\u$, outputs $\y$ and internal variables $\x$.}
    \label{fig:system}
\end{figure}
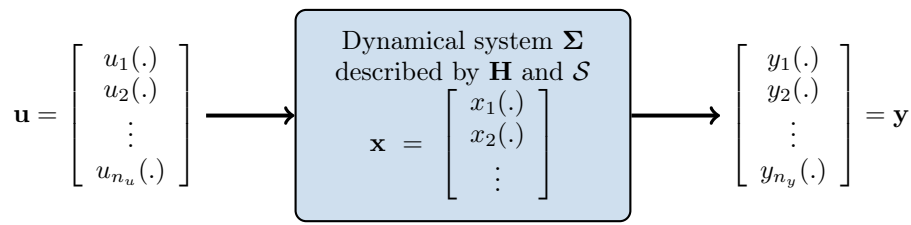

The external, frequency-domain description of $\Hsystem$ is related its internal, time-domain description by the expression of its transfer function $\Htran$ \emph{w.r.t.} the matrices of the realisation $\Hreal$ when the pencil $(\E,\A)$ is regular,
\begin{equation}
  \Htran(s) =  \C\pare{s\E-\A}^{-1}\B.
  \label{eq:ftot}
\end{equation}
Figure \ref{fig:system} provides a graphical view of the system $\Hsystem$, as well as its \emph{(i)} external representation $\Htran$ from $\u$ to $\y$ and \emph{(ii)} its internal view $\Hreal$, including $\x$.

%If the matrix pencil $(\E,\A)$ is regular, $\Htran(s) = \C(s\E-\A)^{-1}\B$, is called  the transfer function associated to the realisation $\Hreal$ of the system $\Hsystem$.

Readers may note that unlike most control textbooks, the internal representation $\eqref{eq:S}$ does not contain any $\D$ term in the output while there is a matrix $\E$ in the dynamical equation. It turns out that the description \eqref{eq:S} can encompass the $\D$ matrix and actually allows to describe a wider class of systems (see \eg \cite{AntoulasBook:2005,AntoulasBook:2020} for interesting discussion on this topic).

%Reader may note that the $\E$ matrix is considered here while no $\D$ term is present (\ie $\y(t)= \C\x(t)$ instead of $\y(t) = \C\x(t)+\D\u(t)$). In the control textbooks the inverse setting is classically considered, but without entering into details, reader should accept that such a formulation is way more complete and allows describing a larger model, and consequently system, class (see \eg \cite{AntoulasBook:2005,AntoulasBook:2020} for interesting comments).

As schematised in Figure \ref{fig:system}, continuous \mimo \lti dynamical model (or system) $\Hsystem$ defines an ''input-output'' map associating an input signal $\u$ to an output one $\y$ by means of the convolution operation,
$$
\y(t)  = \mathbf h(t)*\u(t) = \int_{-\infty}^{\infty} \mathbf h(t-\tau)\u(\tau)d\tau,
$$
where $\mathbf h(t)$ is the impulse response of the system $\Hsystem$.  It is (strictly) causal if and only if $\mathbf h(t)=0$ for ($t\leq0$) $t<0$ (in this case $\E$ is full rank).  Taking the Laplace transform of this input-ouput mapping leads to equation \eqref{eq:H} where $\u(s)$ and $\y(s)$ are the Laplace transform of $\u(t)$ and $\y(t)$, respectively. An \lti system $\Hsystem$ is said to be stable if and only if its transfer function $\Htran$ is bounded and analytic on $\Cplx_+$, \ie it has no singularities on the closed right half-plane. Conversely, it is said to be anti-stable if and only if its transfer function  is bounded and analytic on $\Cplx_-$ (see also \cite{Hoffman:1962} or Chapter 2 of \cite{PontesPhD:2017} for more details)\footnote{At this point, one may admit that different stability notions exist: internal, input-output\dots Once again, this is out of the scope of this report.}. When $\Htran$ is associated with a first order descriptor realisation $\Hreal:(\E,\A,\B,\C)$ as in \eqref{eq:S}, it is rational and has a finite number of singularities called poles or (eigen-)modes of the system. These poles are the singularities of the $(\E,\A)$ pencil $\forall \lambda \in \Cplx$ as $\mathbf{det}(\A-\lambda \E)$. This pencil is regular if at least there exist $\lambda$ such that $\mathbf{det}(\A-\lambda \E)\neq 0$. We call $\lambda$ an eigenvalue of $(\E,\A)$ if $\mathbf{det}(\A-\lambda \E)= 0$.

%In the case where $\Htran$ is rational, it has a finite number of singularities and can be represented by a with $n_u$ inputs, $n_y$ outputs and $n$ internal variables as in \eqref{eq:S}.

Based on this fairly general introduction of variables and elements of linear dynamical systems theory, the remainder of this tutorial is restricted to a simple example.

%Now variables descriptions and fairly general comments on linear dynamical systems and models have been recalled, let us restrict and illustrate the rest of our tutorial using a simple example.

\subsection{Considered use-case}

For didactical purpose, the following assumption will be considered: the system $\Hsystem$ is a \emph{(i)} stable \emph{(ii)} Single Input Single Output (\siso), i.e. $n_u=n_y=1$ and  \emph{(iii)} described by strictly proper and rational function ($\E$ is full rank and thus does not admit a direct feed-through term). Authors stress that while  assumptions \emph{(i)}-\emph{(ii)} are chosen for educational purpose, assumption \emph{(iii)} is also related to more complex issues omitted here and where details may be found \eg in \cite{PoussotHDR:2019,AntoulasBook:2020}.

In this report, a simple yet interesting second order \lti \siso system $\Hsystem$ use-case is considered. Let us assume that the system $\Hsystem$ to be studied is described by the (unknown) following transfer function $\Gtran$ and realisation $\Hreal$,
\begin{equation}
\begin{array}{rcl}
\y(s) &=& \mathbf G(s) \u(s) \\
      &=& \dfrac{k}{s^2/w_0^2 + 2ds/w_0 +1} \u(s) \\
      &=& \underbrace{\vectortwoT{0}{w_0^2}}_{\C}\underbrace{\pare{sI_2-\matrixtwo{-dw_0}{w_0\sqrt{1-d^2}}{-w_0\sqrt{1-d^2}}{-dw_0}}^{-1}}_{\pare{s\E-\A}^{-1}}\underbrace{\vectortwo{k/w_0\sqrt{1-d^2}}{0}}_{\B} ,
\end{array}
\label{eq:G}
\end{equation}
where $d=0.2$ is the damping ratio (note that it is standard to consider $0<d<1$), $\omega_0=\sqrt{\omega_1^2+d^2}$rad/s (where $\omega_1=10$rad/s is the cut-off frequency) and $k$ is the static gain. Such a model represents a simple weakly damped ($d<\sqrt{2}/2$)  second order model system with cut-off dynamic at $\omega_1$. The corresponding Bode diagram, representing the response of $\Gtran(s)$ for varying values of $s=\imath\omega$, where $\omega\in\Real$ (\ie the response of the complex function along the imaginary axis $\imath\Real$) is given in the following Figure \ref{fig:G}, where the gain is shown on the top frame while the phase on the bottom one. Note that the gain exhibits a bump at $f_1=\omega_1/(2\pi)$ due to the low damping $d$ value, a static value above 0dB and a roll-off below $\omega_1$. Regarding the phase, a sharp drop of $-\pi$ at $f_1$ is also observed (note that a $x$-axis log-scale and gain $y$-axis in dB are standardly used).% (due to second order).

\begin{figure}[h]
\centering
\includegraphics[width=.6\columnwidth]{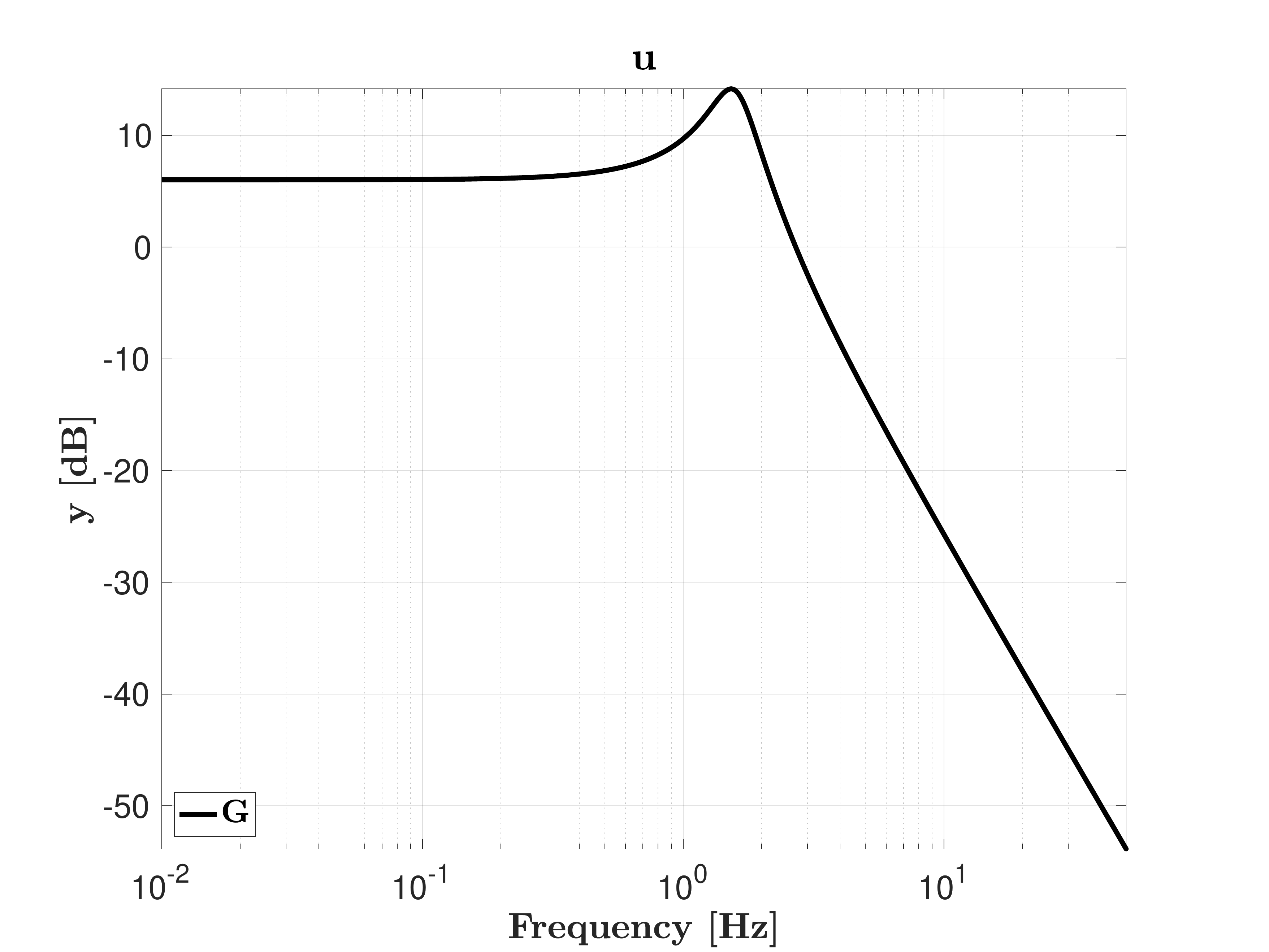}
\includegraphics[width=.6\columnwidth]{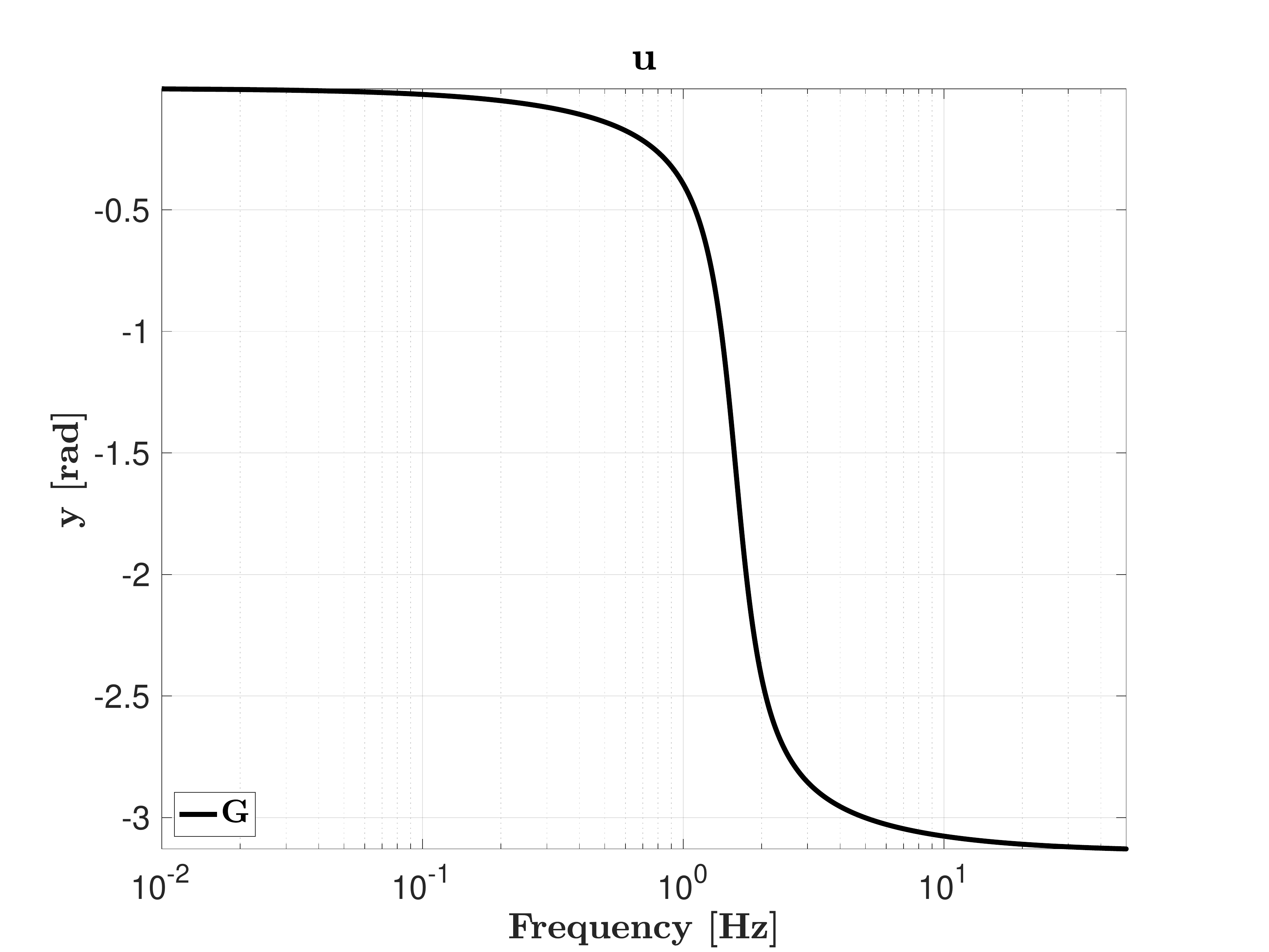}
\caption{Frequency response gain (top) and phase (bottom) of the example model $\Gtran$ (\code{G} in the code).}
\label{fig:G}
\end{figure}

The associated realisation is represented by two first order \ode indicating that two storage states are embedded in the system. Note that the choice of the $(\E,\A,\B,\C)$ quadruple, being constant linear matrices in $(\Real^{n\times n},\Real^{n\times n},\Real^{n\times n_u},\Real^{n_y\times n})$,  has been done among an infinite set of possibilities. Indeed, by considering any full-rank matrix $V \in \Real^{n\times n}$,  the projected realisation  $(V^{-1}\E V,V^{-1} \A V, V^{-1} \B,\C V)$ leads to the same transfer function, and thus input output relation. The following \matlab code provides a way to define such a system.

\begin{lstlisting}[firstline=7,lastline=12,caption={Example file \codeScript{start2ndOrder.m}: $\Gtran$ model description.}]
% Model description as Transfer Function and State-Space
d   = .2;
k   = 2;
w1  = 10; w0 = sqrt(w1^2+d^2);
G   = tf(k,[1/w0^2 2*d/w0 1]); G.InputName = 'u'; G.OutputName = 'y';
Gss = ss([-d*w0 w0*sqrt(1-d^2); -w0*sqrt(1-d^2) -d*w0],[0; w0^2],[k/(w0*sqrt(1-d^2)) 0],0);
\end{lstlisting}

\subsection{Reader target and some advertisement}

As stated in the abstract, this reports tries to cover a wide spectrum of linear signal processing and control theory aspects, and reader should be aware that the present document is not complete at all and should be amended and enhanced in many ways. It is principally meant at providing very basic elements that may be considered as control starting points for non experts people. The report is centred on the engineering soundness and thus focuses on the practical aspects rather than on the theoretical ones, even if authors try paying attention in giving accurate explanations, especially in the transitions from a part to an other.

\subsection{Report structure}

The present report follows the - control engineer-like - classical flow: Section \ref{sec-signals} exposes some elements about the generation of exciting signals, \ie signals that can be used for model construction. Based on these excitation signals and the resulting measured outputs, Section \ref{sec-approximation} then explains how an (approximate) \lti model ($\Htranr$) $\Htran$ of the system $\Hsystem$ can be constructed. Hopefully, ($\Htranr$) $\Htran$ should be identical to $\Gtran$ as defined in \eqref{eq:G}. As it is now well admitted that it is an efficient and versatile tool for linear model approximation and reduction, the lens of the model approximation and interpolatory framework is chosen to be the identification and approximation main tool\footnote{This section voluntarily does not mention explicitly model identification as it is out of the author's knowledge. The model interpolation approach is used instead. Even if authors are convinced that these approaches are really close, obviously this choice can be discussed - over pages and pages - but it is out of the topic of the note.}. Section \ref{sec-control} then describes a model-based continuous-time linear controller $\mathbf K$ design approach, using the $\Hinf$-norm minimisation criteria. Again, such a choice has been done purposely considering the author's background in control theory, but (m)any other approach can be used and be somehow approached with the same philosophy. One major benefit of the $\Hinf$ approach is that it is relatively simple to understand and the available tools for optimising the control structure and gains are powerful, very versatile and accessible to many engineers. This section closes with the discretisation $\mathbf K_z$ of the aforementioned controller and provides preliminary comments on the hybrid interconnection (continuous/sampled). Section \ref{sec-modulation} present an issue encountered by many practitioners, namely the modulation of the control signal to tackle the case where actuators are pulsed (on/off). This modulation choice is done considering the numerous discussions authors got from practitioners and therefore seems of interest. As it is a fairly complex point, this last section is mainly based on simulations and the discussion remains preliminary. Conclusions and some questions are discussed in Section  \ref{sec-conclusion}.

\newpage
\section{Identification signals}
\label{sec-signals}
%\subsection{General idea}

The main aim of identification signals is to generate an input signal exciting enough to enable an accurate identification of the underlying system, here represented by a linear second order model $\Gtran$. Such a signal should then satisfy some hypothesis such as applicability, be exciting enough and simple. In general for linear systems, the frequencies of identification signal should excite all the relevant frequencies of the system. Once again, as it is quite far from the main expertise of the authors, reader should take the following section as a naive, but simple to reproduce, approach.

\subsection{Reminders on Fourier transform}

As most of the report is based on the frequency-domain representation of dynamical systems, let us first recall some basic elements on Fourier transform. Given signal $\x(t)$ the Fourier transform reads
\begin{equation}
\begin{array}{rcl}
    \tilde \x(f) = \tf{\x(t)} &=& \displaystyle\int_\Real \x(t)e^{-2\pi \imath ft}dt \\
    &=& \abs{\tilde \x(f)} e^{\imath \phi(f)},
\end{array}
\end{equation}
where the complex variable $\tilde \x(f)$ is the spectrum of $\x(t)$ and where  $\phi(f)$ denotes the argument of $\tilde \x(f)$. The time $t$ and frequency $f$ are the variables embedded in  the transform. Importantly, the Fourier transform exists for any finite energy signals $\x(t)$ (it is a sufficient condition). One may note that the dual inverse Fourier operator exists. As a remark, one can remember that signals with small support has a large spectrum. As an illustration, given the rectangle function (note tat such shape has some interest in application, as shown at the end of this report)
\begin{equation}
    \x(t) = \textbf{rect}_T(t/T) \left\{
    \begin{array}{ll}
    1 & \text{, $\forall t \in [-T/2,T/2]$} \\
    0 & \text{, elsewhere}
    \end{array}
    \right.
\end{equation}
the corresponding Fourier transform reads
\begin{equation}
    \tilde \x(f) = \dfrac{\sin(\pi f T)}{\pi f} = T \mathbf{sinc}(\pi T f).
\end{equation}
%being wider when $$

\subsection{Some exciting signal}

Back to our problem, the first objective of a control engineer is to construct exciting signal that will allow us for constructing a linear model $\Htran$ of the system $\Hsystem$, here being $\Gtran$ but considered as unknown. There exists a lot of identification signals and it is actually a complete research field in systems theory, but for simplicity, among them, one can mention the (fairly standard) following ones:
\begin{itemize}
    \item Pseudo random binary signal (\textbf{PRBS}). It consists in on/off like signals used as input. The idea is to define a "random" sequence of these signals in order to emulate a white noise. The longest duration should last enough to reach steady state output and the shortest should be short enough to excite frequencies above the cut-off one of the system. The Fourier transform of such signal should be constant over frequencies. Note that for the same reason as the one mentioned above in the rectangular signal, to be a white (more accurately pink) noise, a whitening filter should be combined to the random rectangular sequence, avoiding the frequency-domain zeros embedded in the $\mathbf{sinc}$ function.
    \item Frequency chirp signal. It consists in a  cosine-like function that sweeps from a low frequency up to a high one. The signal should sweep sufficiently slowly to excite all frequencies. Similarly to the above random binary signal, the Fourier transform covers all frequencies in the interval of the sweep. An illustration of such a signal and its response when fed in the example use case $\Gtran$ is given in Figure \ref{fig:chirp}.

    \begin{figure}[H]
    \centering
    \includegraphics[width=.8\columnwidth]{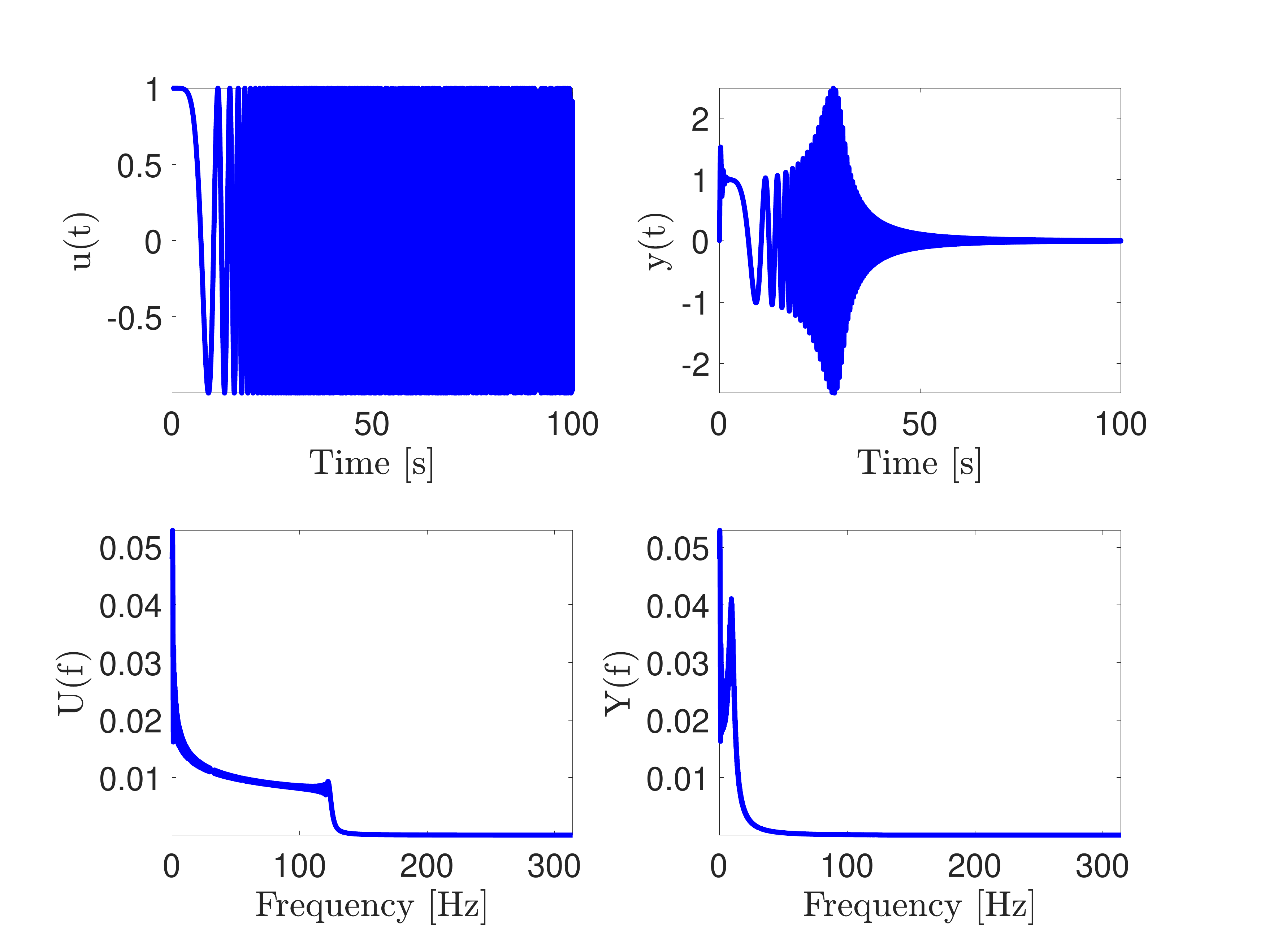}
    \caption{Chirp (frequency sweep) signal. Top: time-domain input signal $\u(t)$ (left) and output measurement $\y(t)$ (right). Bottom: frequency-domain corresponding signals $\tilde \u(f)$ and $\tilde \y(f)$, respectively.}
    \label{fig:chirp}
    \end{figure}

    \item Impulse signal. It consists in injecting a causal Dirac input, denoted $\delta(t)$. Theoretically, a Dirac signal reads
    \begin{equation}
        \delta(t) = \left\{
        \begin{array}{ll}
            1 & \text{, $t=0$} \\
            0 & \text{, elsewhere}
        \end{array}
        \right. .
    \end{equation}
    It naturally translates as
    \begin{equation}
       \tilde \x(t) = \tf{\x(t)} =  \int_{-\infty}^{+\infty} \delta(t) dt = 1.
    \end{equation}
    The Fourier transform of the Dirac function is a unitary gain over all frequencies. Thus, Dirac functions encompasses all frequencies (note that a continuous signal contains the frequency zero, only). In practical signal processing, it is used for frequency modulation. In practice, such a signal is impossible to generate and a rectangular signal with small width is preferred. The resulting spectrum is generally a cardinal sine like and thus also a pink noise. An illustration of such a signal and its response when fed in the example use case $\Gtran$ is given in Figure \ref{fig:impulse}.
    \begin{figure}[H]
    \centering
    \includegraphics[width=.8\columnwidth]{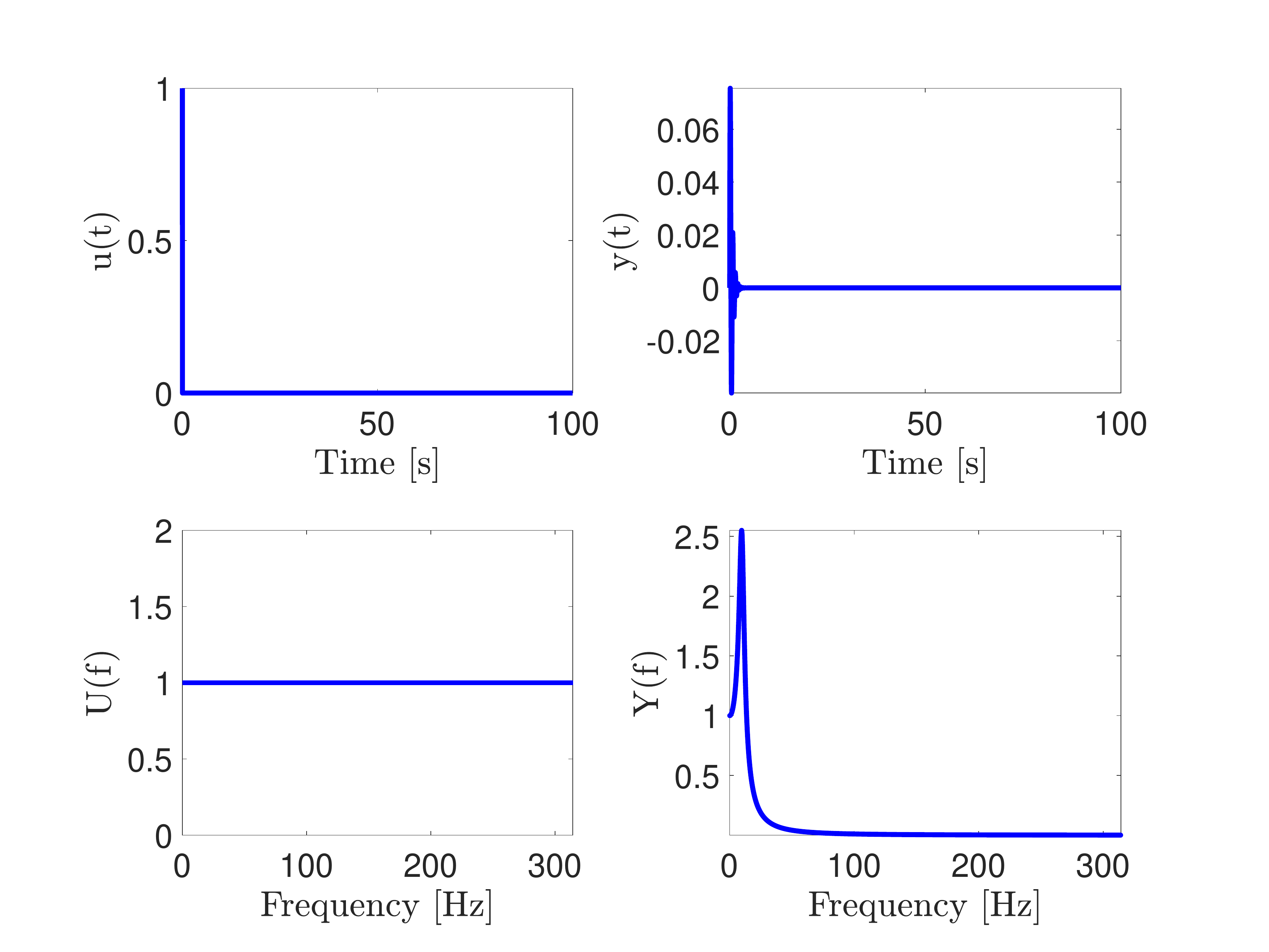}
    \caption{Impulse signal. Top: time-domain input signal $\u(t)$ (left) and output measurement $\y(t)$ (right). Bottom: frequency-domain corresponding signals $\tilde \u(f)$ and $\tilde \y(f)$, respectively.}
    \label{fig:impulse}
    \end{figure}
\end{itemize}

Note that as we consider linear systems only, one single amplitude is enough for identification. Obviously, in nonlinear cases, this statement is not true anymore. From the rest of the report, we will consider results obtained when using the second exciting signal, \ie the frequency sweep one. The signals are obtained using the following \matlab code with \code{inputExcitation='chirp'}.
\begin{lstlisting}[firstline=14,lastline=41,caption={Example file \codeScript{start2ndOrder.m}: generate and simulate a chirp signal excitation.}]
switch inputExcitation % here = 'chirp'
    case 'chirp'
        Tsamp = 1e-2;
        Tf    = 1e2;
        T     = 0:Tsamp:Tf; % time sample
        F0    = 1e-6;       % lower chirp frequency
        F1    = 2/Tsamp/10; % higher chirp frequency
        T1    = T(end);
        Uchi  = mor_chirp(T,F0,T1,F1,'quadratic');
        U     = Uchi;
    case 'impulse'
        Tsamp = 1e-2;
        Tf    = 1e1;
        T     = 0:Tsamp:Tf;
        Uimp  = mor_impulse(T); 
        U     = Uimp;
    case 'prbs'
        Tsamp = 1e-1;
        NN    = 10;
        Trep  = 5;
        Tseq  = Trep*(2^NN-1);
        Tf    = 1*Tseq;
        T     = 0:Tsamp:Tf;
        M     = floor(Trep/Tsamp/NN);
        Uprbs = mor_prbs(T,NN,M);
        U     = Uprbs;
end
Y = lsim(G,U,T); % simulate the system G with U, over T
\end{lstlisting}

\subsection{Transfer estimation}

Based on the input $\u(t)$ and output $\y(t)$ signals and on their frequency equivalence  $\tilde \u(f)$ and $\tilde \y(f)$, the cross correlation transfer can be computed. In opposition to the spectral energy density, the cross (or inter-correlated) spectral density is a complex number which gain represents the interaction power and which arguments represents the phase between $\tilde \u(f)$ and $\tilde \y(f)$. A simplified \matlab code allowing to obtain such frequency response is given in what follows.

\begin{lstlisting}[firstline=43,lastline=56,caption={Example file \codeScript{start2ndOrder.m}: a simplified frequency transfer response computation.}]
% Frequency transfer estimation from U to Y
Fsamp    = 1/Tsamp;     % frequency sampling
Fnyq     = Fsamp/2;     % Nyquist frequency
NFFT     = [];
L        = numel(U);
FTy      = fft(Y,NFFT); % Fourier transform of y
FTu      = fft(U,NFFT); % Fourier transform of u
F        = linspace(0, 1, fix(L/2)+1)*Fnyq;
Iv       = 1:numel(F);
FTy      = FTy(Iv);
FTu      = FTu(Iv);
Txy      = FTy./FTu;    % discrete transfer estimation
H(1,1,:) = Txy;
W        = 2*pi*F;      % pulsation points
\end{lstlisting}

\begin{figure}[h]
\centering
\includegraphics[width=.6\columnwidth]{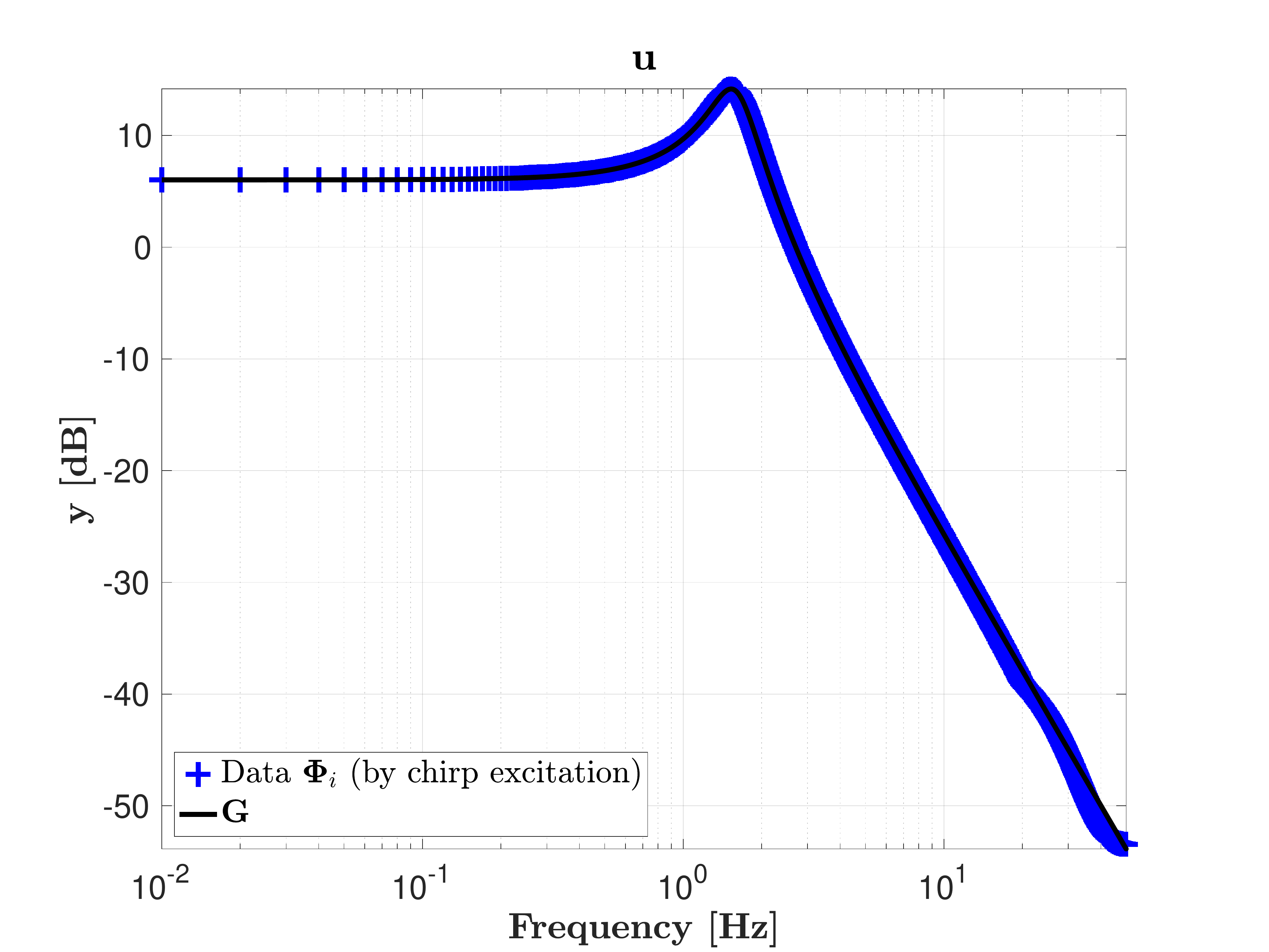}
\includegraphics[width=.6\columnwidth]{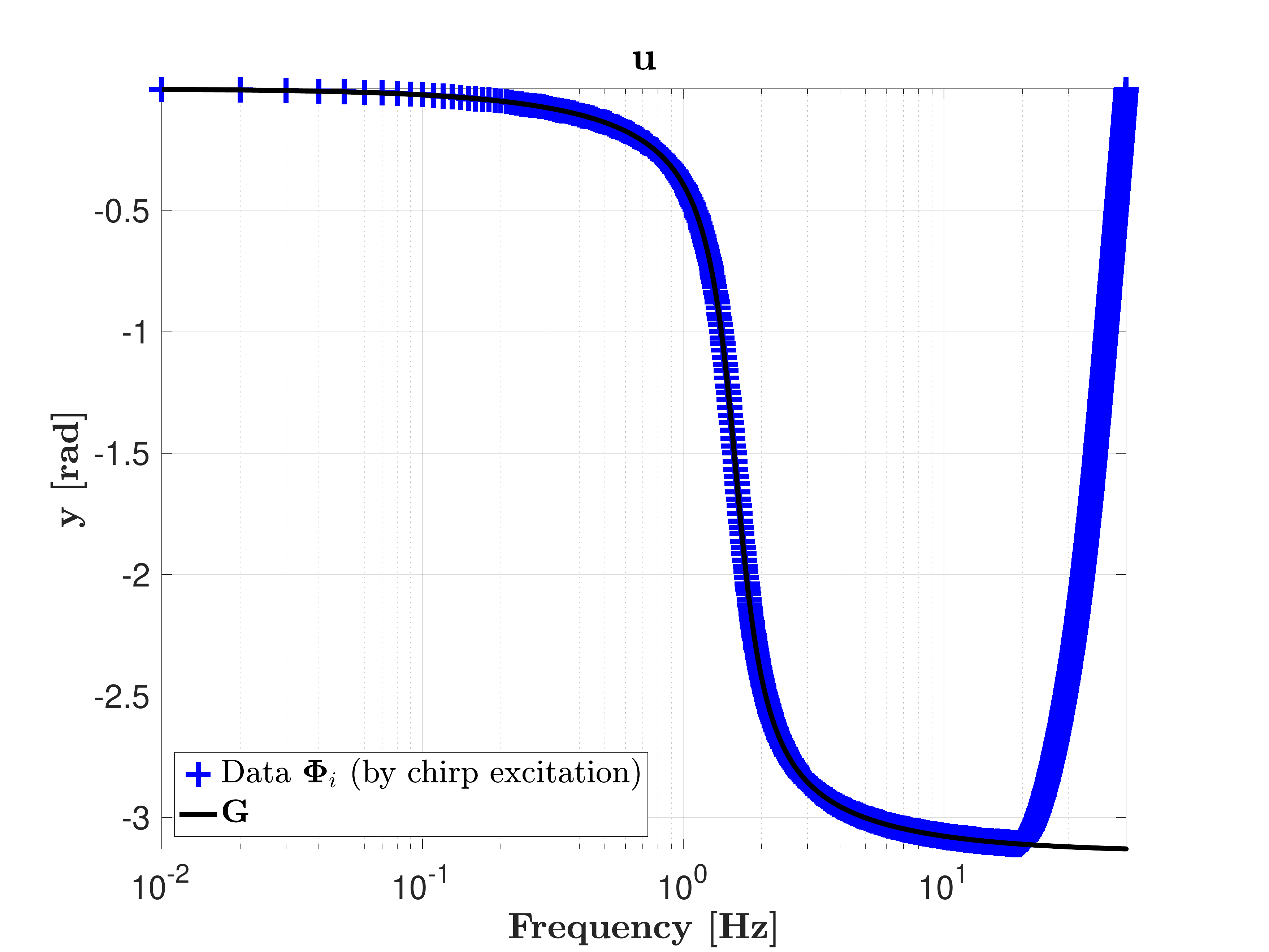}
\caption{Frequency response gain (top) and phase (bottom) of $\Gtran$ (\code{G}) and its discrete estimation $\{\omega_i,\mathbf \Phi_i\}$ ($\code{\{W,H\}}$) using the chirp excitation signal.}
\label{fig:freqresp}
\end{figure}

When applied on the chirp signal case presented above, the above discrete frequency response estimation $\mathbf \Phi_i=\y(\imath\omega_i)/\u(\imath\omega_i)\in\Cplx^{n_y\times n_u}$ (denoted \code{H}), estimated at samples $\imath\omega_i=\imath 2\pi f_i$ (\code{W}), where $i=1,\dots,N$, is obtained and leads to Figure \ref{fig:freqresp}. Interestingly, when comparing the the original linear system $\Gtran$, the overall dynamic seems well cached, even if some errors appear in high frequencies (close to the Nyquist frequency). This observation is consistent with the fact that the chirp function is exciting enough. Note that the frequency plot stops at the Nyquist frequency  $f_{\text{Nyquist}}=f_s/2=50$Hz, where $f_s=100$Hz. Due to the sampling effect, above this Nyquist, the spectrum repeats with this frequency periodicity.

By analysing the frequency response of $\Gtran$ and its estimated discrete samples $\mathbf \Phi_i$, some differences are observed, even with no noise considered in the measurements. Signal processing theory can bring answers to this discrepancies by considering the effect of sampling and reducing it by using more elaborate techniques. However, at this point, the transfer estimation seems acceptable. Note that in practice, an exact model appears to be a unicorn as model validity can only be considered through the lens of (experimental) data which are affected by many uncertainties. A model should be considered mainly \emph{w.r.t.} what it will be used for. In our case, the model will mostly serve at designing a controller, and variability should always be considered. This remark is also linked with what people call robustness with respect to some model variability. This specific point is not directly addressed here but interested reader may find clues and more detailed theorems in the very complete book \cite{ZhouBook:1997}.

\newpage
\section{Reduced model construction}
\label{sec-approximation}
As rooted on the input-output collected data from the exciting signal followed by the Fourier transform and cross correlation transfer estimation, one now has access to the $\{\imath\omega_i,\mathbf{\Phi}_i\}_{i=1}^N$ couple, denoted $\code{\{W,H\}}$. From this basis, we are now ready to perform the model construction step. Many techniques are tailored to this objective, but here, let us invoke the interpolatory framework to construct a rational model $\Htran$ as well as its approximation $\Htranr$ (see \cite{AntoulasBook:2020} for complete description and presentation of "interpolatory" problem and meaning).

\subsection{Data-driven approximation}

Given the complex-valued input-output data collection $\{z_i,\mathbf{\Phi}_i\}_{i=1}^N$ or more specifically in our case $\{\imath\omega_i,\mathbf{\Phi}_i\}_{i=1}^N$  (where $z_i\in\Cplx$, $\omega_i\in\Real$ and $\mathbf \Phi_i\in \Cplx^{n_y\times n_u}$) defined as,
\begin{eq}
\y(z_i) = \mathbf \Phi_i \u(z_i) \text{  or  } \y(\imath\omega_i) = \mathbf \Phi_i \u(\imath\omega_i),
\label{eq:data}
\end{eq}
the approximation problem aims at constructing the approximate rational transfer function matrix $\Htranr$ mapping inputs $\u$ to the approximate outputs $\yr$ such that
\begin{eq}
\yr(s) = \Htranr(s) \u(s).
\label{eq:transferRed}
\end{eq}
Obviously, some objective are that \emph{(i)} the reduced inputs to outputs map should be "close" to the original \ie for the same $\u$, $\yr$ close to $\y$ in some sense, \emph{(ii)} the critical system features and structure should be preserved, and,  \emph{(iii)} the strategies for computing the reduced system should be numerically robust and stable. Approximating $\Gtran$ with \eqref{eq:transferRed} is a {model-based} approximation, while, approximating its input-output data $\{\imath\omega_i,\mathbf{\Phi}_i\}_{i=1}^N$ with \eqref{eq:transferRed} belongs to the {data-driven} family (see \cite{AntoulasBook:2005,PoussotHDR:2019} for examples). Here we first follow the {data-driven} philosophy and secondly the {model-based} one. In both cases, the interpolation lens of is used.

In the data-driven model approximation, the main ingredient is the Loewner framework initially settled in \cite{Mayo:2007}. Interested reader  can  also find details in \cite{AntoulasBook:2020} and practical clues and applications in \cite{PoussotHDR:2019}. In brief, the Loewner approach is a data-driven method building a rational descriptor \lti dynamical model $\Htran$ of dimension $m$ of the same form as \eqref{eq:S}, which interpolates frequency-domain data given as \eqref{eq:data}. It is rooted on the so-called Loewner and shifted Loewner matrices which provide information on the minimal order of the interpolating underlying rational function. The \mor toolbox provides an implementation of this method, which can be called as follows

\begin{lstlisting}[firstline=87,lastline=94,caption={Example file \codeScript{start2ndOrder.m}: data-driven model construction.}]
% Approximation from data
opt            = [];
opt.verbose    = true;
opt.sample     = floor(length(W)/256);  % undersampling to fasten approximation
opt.freqBand   = [0 Fnyq/3]*2*pi;       % frequency band of approximation
opt.ensureStab = true;                  % enforce model stability
[Hi,info_i]    = mor.lti({W,H},[],opt); % [] means that the exact order model is sought
\end{lstlisting}

On the basis of the frequency \code{W} (\ie $\imath\omega_i$) and corresponding frequency responses \code{H} (\ie $\mathbf \Phi_i$) set, the above code computes \code{Hi} (\ie $\Htran$), a minimal order interpolating rational function equipped with a descriptor state-space realisation. With the arguments provided in the \code{mor.lti} routine, the approximation is done up to the frequency $f_{\text{Nyquist}}/3$ (indeed when analysing the data obtained during the transfer function estimation, strange behaviour is observed after this frequency and are thus discarded in this step), data are under-sampled (for numerical simplicity) and the output \code{Hi} model is forced to be input-output stable. The following information are also listed.
\begin{lstlisting}
+------------------------------------------------------------------------------+
| MOR Toolbox                                                                  |
| Loewner - Loewner Interpolation Algorithm                                    |
+------------------------------------------------------------------------------+
| Right data : {la_i,r_i  ,H(la_i)  =w_i  } i = 1...k                          |
| Left data  : {mu_i,l_i^T,H(mu_i)^T=v_j^T} j = 1...q                          |
|                                                                              |
| Loewner matrix real form is computed from complex data                       |
| All rank cond. not checked      : consider directions change                 |
| Rational function dimension (n) : 68                                         |
| Mc Millian degree (nu)          : 59                                         |
| Minimal realization degree (r)  : 58                                         |
| r~=nu                           : D or polynomial case                       |
| Selected order                  : 58                                         |
| RHinf sigma/gamma : 0.3587/0.3587 (optimal)                                  |
+------------------------------------------------------------------------------+
\end{lstlisting}

Briefly, the above information set displayed after executing this \mor toolbox code mainly indicates that an order $n=58$ of function $\Htran$ (\code{Hi}) has been obtained. This order is automatically computed by the procedure. At this point, a model $\Htran$ of dimension $n=58$ is then obtained. Of course, considering the original system $\Gtran$ (of oder 2), such an order is way too large, but considering the interpolatory objective this not specifically strange. We won't enter into details here and will focus our attention on its frequency response, later presented on Figure \ref{fig:freqresp}, showing that it really well capture the data collected, and restitute the behaviour of $\Gtran$ (obviously with a way too large function). As this function is of high order, in view of control design, it is well admitted that a reduction step is necessary. Indeed, most of the control methods are very limited to model with low order models and numerical accuracy is expected with simpler models.

\subsection{Model reduction}

As rooted on the obtained interpolated model $\Htran$ (\code{Hi}), a $n_u$ inputs, $n_y$ outputs linear dynamical system  described by the complex-valued function from $\u$ to $\y$, of order $n$ ($n$ large or $\infty$)
$$
\Htran:\Cplx \rightarrow \Cplx^{n_y\times n_u},
$$
the model approximation problem consists in finding $\Htranr$ of order $r\ll n$
$$
\Htranr:\Cplx \rightarrow \Cplx^{n_y\times n_u},
$$
that well reproduces the input-output behaviour and equipped with realisation
$$
\Htranr(s) = \Cr(s\Er  - \Ar)^{-1}\Br,
$$
where $\Er,\Ar \in \mathbb{R}^{r \times r}$, $\Br \in \mathbb{R}^{r\times n_u}$ and $\Cr \in \mathbb{R}^{n_y \times nr}$, are constant matrices. One standard way to deal with this problem is to consider the $\Htwo$ model approximation problem given as follows.
$$
\Htranr := \arg \min_{
\small
\begin{array}{c}
\Gtran\in\Htwo \\
\rank(\Gtran)=r \ll n
\end{array}
\normalsize
} \norm{\Htran-\Gtran}_{\Htwo}
$$

Such a problem can be solved using the \mor toolbox through the \code{mor.lti} interface, as detailed in the following code, where one aims at finding an optimal model $\Htranr$ (\code{Hr}) of dimension $r=2$, on the basis of $\Htran$ (\code{Hi}). Here the order two has been selected for illustration purpose, but as $\Htran$  (\code{Hi}) is of dimension $n=58$, different choice may also have been done. Still, as the considered example $\Gtran$ is a second order, let us continue with this assumption.
\begin{lstlisting}[firstline=95,lastline=96,caption={Example file \codeScript{start2ndOrder.m}: model reduction step.}]
% Model order reduction
[Hr,info_r]    = mor.lti(Hi,2,opt);
\end{lstlisting}

Applied on the obtained $\Htran$ (\code{Hi}) model, the above code leads to the following output, illustrating among other, the iterative aspect of the approach.

\begin{lstlisting}
+------------------------------------------------------------------------------+
| MOR Toolbox                                                                  |
| ITIA - Iterative Tangential Interpolation Algorithm                          |
+------------------------------------------------------------------------------+
| Original system      : 58 states, 1 input(s), 1 output(s)                    |
| Reduced system order : 2                                                     |
| H2(W) norm error     : not checked                                           |
| Frequency bound      : [0 104.72] rad/s                                      |
| Shift selection      : automatic                                             |
| Start: 1 /1                                                                  |
|         Iteration             Unconv. shifts               Delta Hr          |
|              1                       2                         -             |
|              2                       2                      151.63           |
|              3                       0                       0.14            |
+------------------------------------------------------------------------------+
\end{lstlisting}

After convergence, the obtained models $\Htran$ (\code{Hi}) and $\Htranr$ (\code{Hr}) Bode responses are shown on Figure \ref{fig:freqresp}. Clearly, it illustrates that the model $\Gtran$ is well captured by $\Htran$ of dimension 58, but also by $\Htranr$ of order 2, being the same order as the original $\Gtran$.
\begin{figure}[h]
\centering
\includegraphics[width=.6\columnwidth]{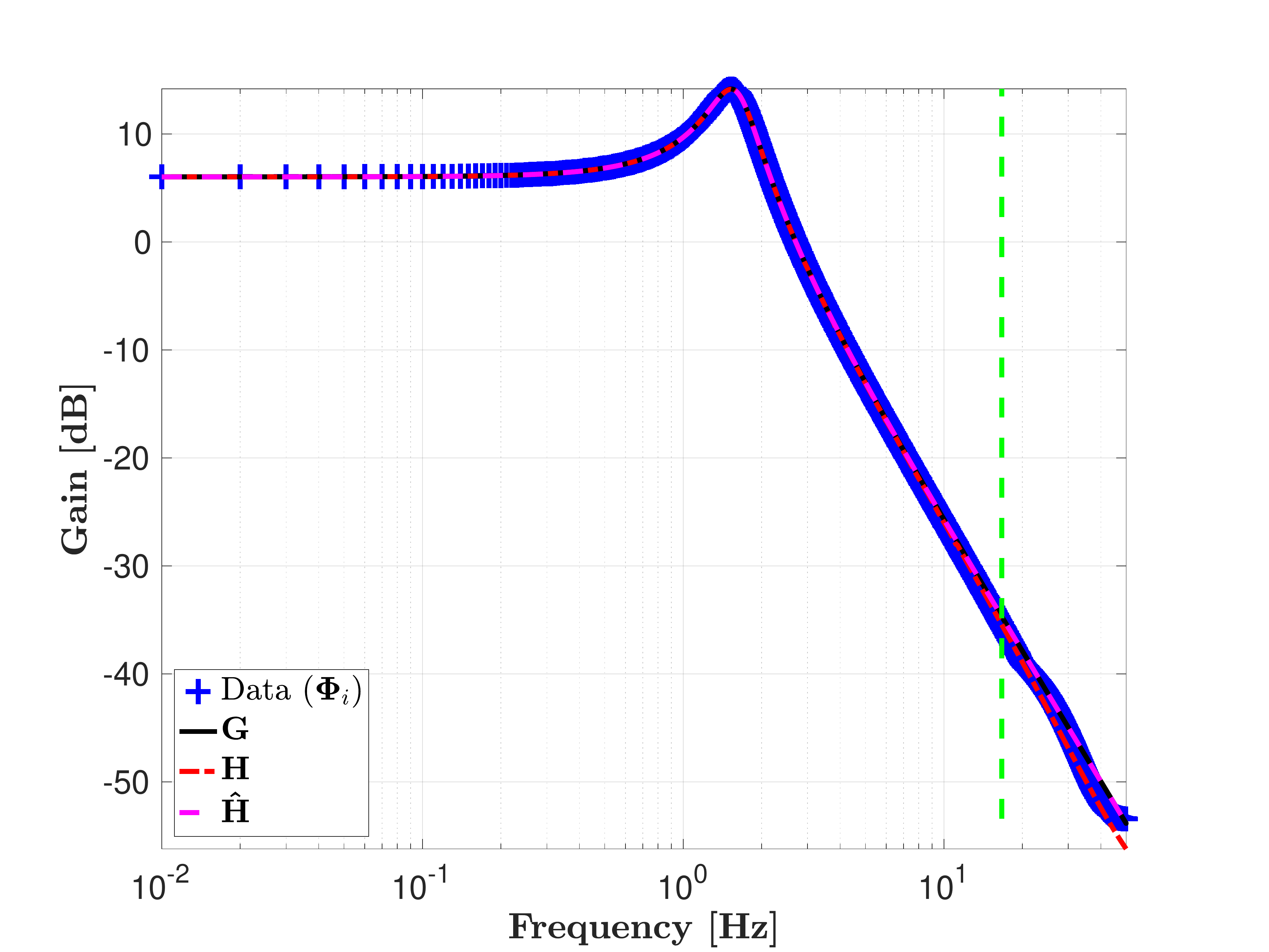}
\includegraphics[width=.6\columnwidth]{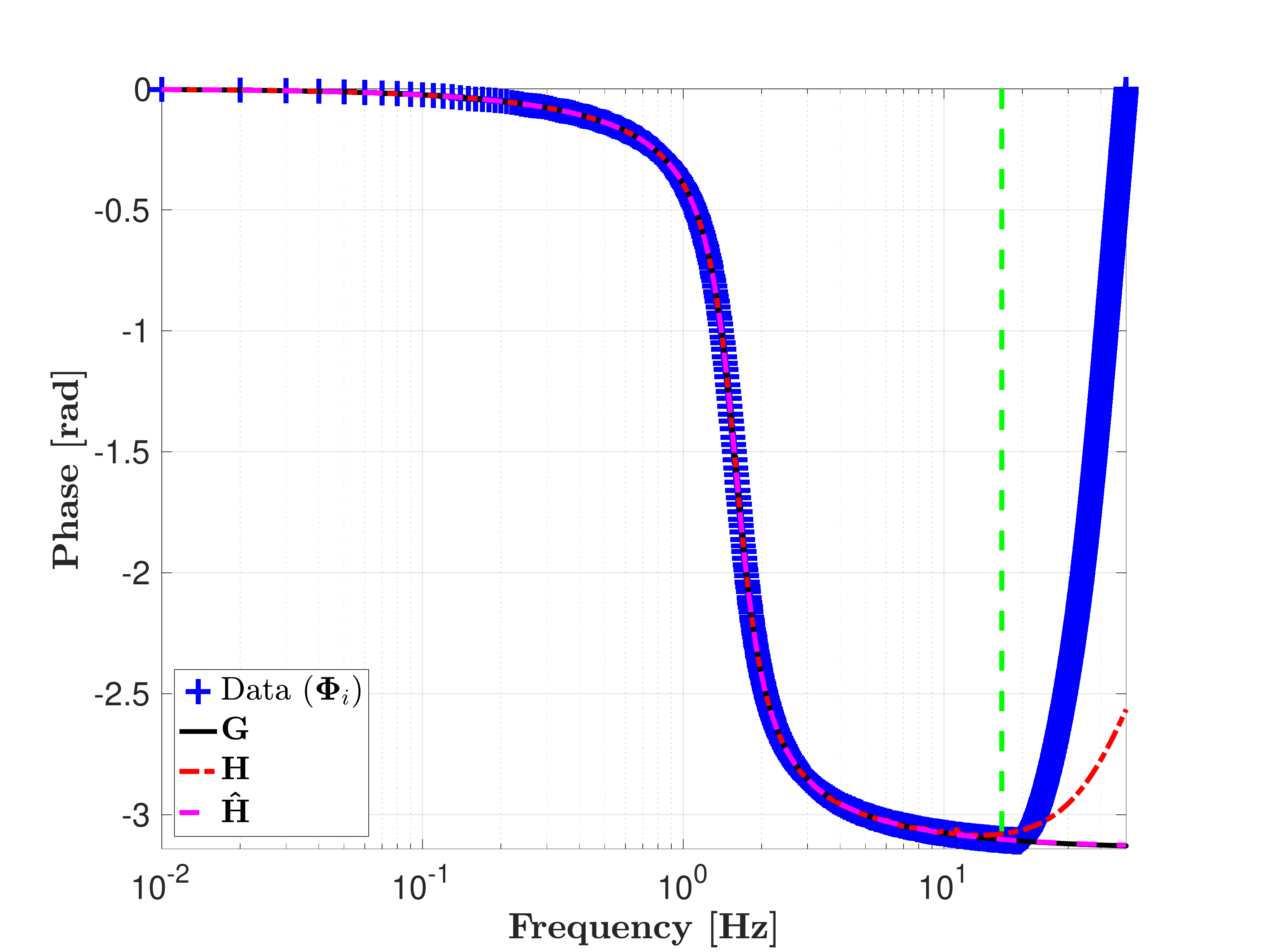}
\caption{Frequency response gain (top) and phase (bottom) of the original data (blue $\mathbf{+}$), original model $\Gtran$ (solid black), full order interpolated model $\Htran$  (\code{Hi} dashed red) and reduced order model $\Htranr$ (\code{Hr} dashed pink).}
\label{fig:freqrespAll}
\end{figure}

\begin{figure}[h]
\centering
\includegraphics[width=.6\columnwidth]{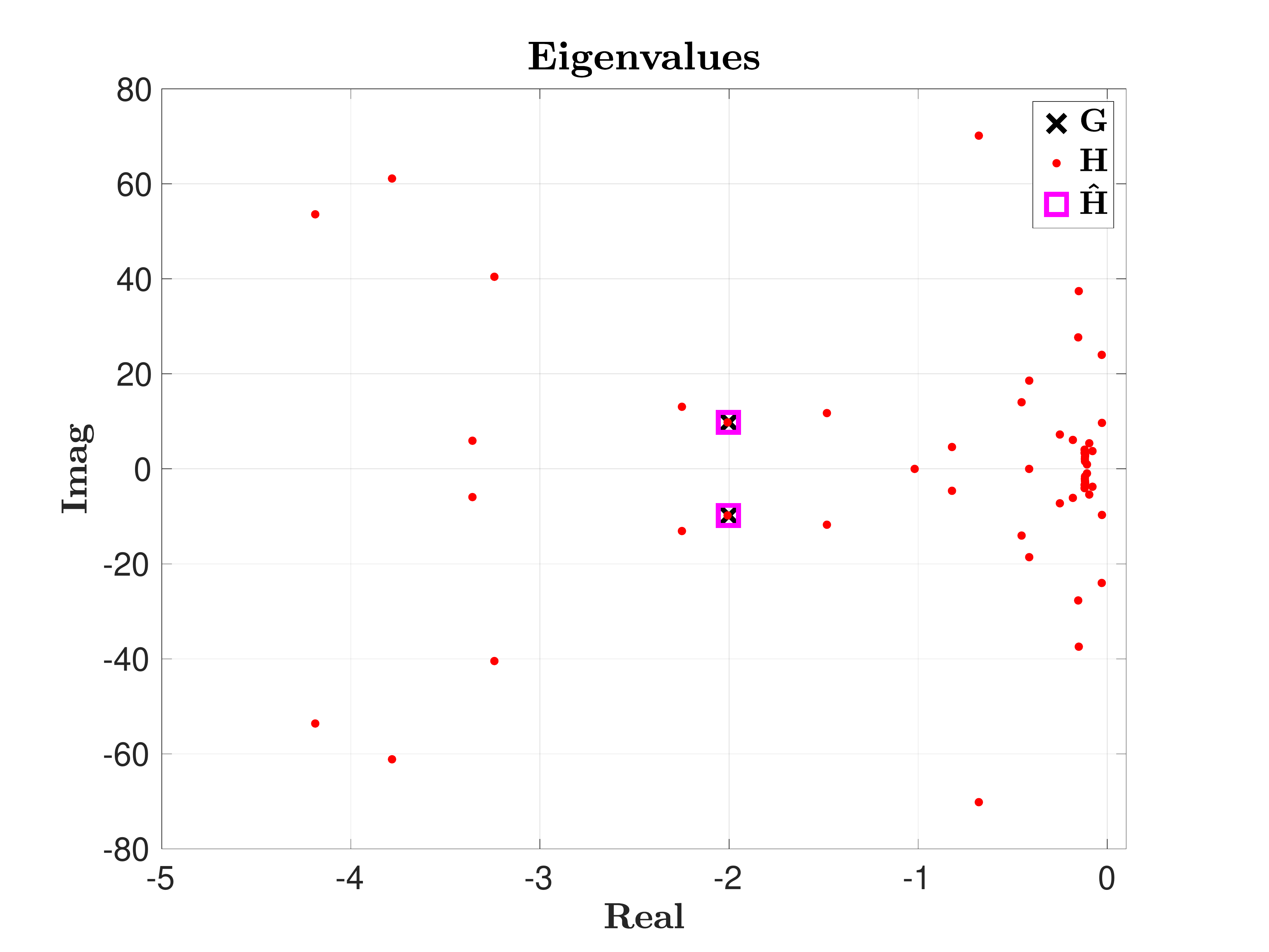}
\caption{Eigenvalues of the original model $\Gtran(s)$ (black $\mathbf \times$), full order interpolated model $\Htran$ (red $\mathbf \cdot$) and reduced order model $\Htranr$ (pink square).}
\label{fig:eig}
\end{figure}

One interesting complement concerns the pencil  associated to the dynamical matrices couple for all models, $\Gtran$, $\Htran$ and $\Htranr$. While the interpolating full order model $\Htran$ exhibits 58 singularities, both the $\Gtran$ ad $\Htranr$ have only two of them. Even more interestingly, these two last models have the exact same eigenvalues, as shown on Figure \ref{fig:eig}, meaning that on the sole basis of system $\Hsystem$ excitation plus model interpolation and approximation, one is able to recover the original model dynamical information without knowing it a priori. This last statement is a very strong one and is obviously valid in linear dynamical systems theory only. It is one of the real strength of model approximation in the linear domain.

\newpage
\section{Control design ($\Hinf$-norm oriented)}
\label{sec-control}
On the basis of the simplified model obtained $\Htranr$ (\code{Hr}), being as accurate as possible but also as simple as possible, we are now ready to design a controller to achieve some closed-loop performances. Here, our simple objective is to make our output $\y$ track an exogenous reference signal denoted $\mathbf r$. Such an objective is a fairly standard one and our aim is more to illustrate in practice how this can be easily done using existing numerical tools. Indeed, authors believe that extension of this problem to more complex cases can be "easily" done once this one well mastered.

\subsection{Preliminary words}

The control design for linear systems is a wide problem on which many researchers and practitioners have proposed methodological results and numerical schemes to achieve different objectives. In this report we only focus on the so called $\Hinf$ control approach, well known in the robust control community for its fantastic robustness and performance definition versatility (see \eg \cite{FrancisBook:1987,ZhouBook:1997} for a very good starting point). Here we follow the $\Hinf$ philosophy which objective is, on the basis of the simplified model $\Htranr$, to design a controller $\mathbf K$ such that,
\begin{equation}
\mathbf K := \arg \min_{
\small
\begin{array}{c}
\mathbf{\tilde K}\in\Hinf \\
\mathbf{\tilde K}\in \mathcal K
\end{array}
\normalsize
} \norm{\mathcal F_l(\Htranr,\mathbf{\tilde K})}_{\Hinf}
\label{eq:hinfPb1}
\end{equation}
where $\mathcal F_l(\cdot,\cdot)$ is the lower fractional operator (see \cite{ZhouBook:1997,MagniLFR:2006} for a good insight), $\mathcal K$ is the class of controller considered (we will come back to this later in this section) and $\Hinf$ denotes either the space of complex-valued functions with bounded supremum over the imaginary axis or the norm associated. More specifically, $\mathbf T_{\mathbf r\y}(\Htranr)=\eval{\mathcal F_l(\Htranr,\mathbf{\tilde K})}{\mathbf{\tilde K}=\mathbf K}$ is nothing but the closed-loop illustrated on Figure \ref{fig:closedLoopContinuous}, when the controller is defined as
$$
\u(s) = \mathbf K(s) \mathbf e(s) = \mathbf K(s) (\mathbf r(s)-\y(s)),
$$
where $ \mathbf e$ is the error signal  and $ \mathbf r$, the reference one.
\begin{figure}[h]
    \centering
    \scalebox{.8}{\tikzstyle{block} = [draw, thick,fill=bleuONERA!20, rectangle, minimum height=3em, minimum width=6em,rounded corners]
\tikzstyle{block2} = [thick,rectangle, minimum height=3em, minimum width=6em,rounded corners]
\tikzstyle{sum} = [draw, thick,fill=bleuONERA!20, circle, node distance=1cm]
\tikzstyle{input} = [coordinate]
\tikzstyle{output} = [coordinate]
\tikzstyle{pinstyle} = [pin edge={to-,thick,black}]
\tikzstyle{connector} = [->,thick]

% The block diagram code is probably more verbose than necessary
\begin{tikzpicture}[auto, node distance=2cm,>=latex']
    % We start by placing the blocks
    \node [input, name=input] {};
    \node [sum, right of=input] (sum) {};
    \node [block, right of=sum, node distance=3cm] (controller) {$\mathbf K(s)$};
    %\node [pinstyle, above of=controller, node distance=1.2cm] (samplingK) {$f_2$};
    %\draw [connector] (samplingK) -- node[name=hK] {} (controller.90);
    %\node [block, right of=controller, node distance=4cm] (pwm) {\textbf{Modulation}};
    %\node [pinstyle, above of=pwm, node distance=1.2cm] (samplingPWM) {$f_1=N f_2$};
    %\draw [connector] (samplingPWM) -- node[name=hPW] {} (pwm.90);
    \node [block, right of=controller, node distance=4cm] (system) {$\Htranr(s)$};
    %\node [pinstyle, above of=system, node distance=1.2cm] (samplingSystem) {$f_1=N f_2$};
    %\draw [connector] (samplingSystem) -- node[name=hPW] {} (system.90);
    \draw [connector] (controller) -- node[name=u] {${\u(t)}$} (system);
    \node [output, right of=system, node distance=3cm] (output) {};

    % Once the nodes are placed, connecting them is easy. 
    \draw [connector] (input) -- node {$\mathbf r(t)$} (sum);
    \draw [connector] (sum) -- node {$\mathbf e(t)$} (controller);
    %\draw [connector] (pwm) -- node {\red{$\mathbf u(t_{k/N})$}} (system);
    \draw [connector] (system) -- node [name=y] {${\y(t)}$} (output);
    \draw [connector] (output)+(-1cm,0) -- ++(-1cm,-2cm) -| node [near start] {} (sum.south);
\end{tikzpicture}}
    \caption{Closed-loop scheme of $\mathbf T_{\mathbf r\y}(\Htranr,\mathbf K)$, being the interconnection of $\mathbf K$ with model $\Htranr$, involved in the synthesis step.}
    \label{fig:closedLoopContinuous}
\end{figure}

Remembering that for \siso systems the $\Hinf$ norm is the peak value of the Bode gain response, as it problem \eqref{eq:hinfPb1} is not really interesting to solve. Indeed minimising the gain of the interconnection shown in Figure \ref{fig:closedLoopContinuous} is not specifically relevant. This why instead of solving \eqref{eq:hinfPb1}, one aims at solving a modified version of it, through what is generally called, the generalised problem, involving the generalised plant. Such a notion is more detailed in the sequel.

\subsection{The $\Hinf$-norm oriented control design}

One essential ingredient in linear control, and so it is in $\Hinf$ control, is the generalised plant concept. Basically, it consists in constructing a plant including the model $\Htranr$ and a set of performance output $\mathbf W_o$ and input $\mathbf W_i$ weighting functions. These weighing functions are interconnected to the plant's model and constitute the basis of the control optimisation process. Practically, they shape the interconnection to be minimised where performances and control objectives are encapsulated in these weighting functions (again, reader may refer to \cite{ZhouBook:1997}). Before illustrating such a generalised plant and how it is constructed, let us just recast the original $\Hinf$ control problem \eqref{eq:hinfPb1} now as
\begin{equation}
\mathbf K := \arg \min_{
\small
\begin{array}{c}
\mathbf{\tilde K}\in\Hinf \\
\mathbf{\tilde K}\in \mathcal K
\end{array}
\normalsize
} \norm{\overbrace{\mathbf W_o\mathcal F_l(\Htranr,\mathbf{\tilde K})\mathbf W_i}^{\mathbf T(\Htranr,\mathbf{\tilde K})=\mathcal F_l(\mathbf P,\mathbf{\tilde K})}}_{\Hinf}
\label{eq:hinfGene}
\end{equation}
where $\mathbf W_o$ and $\mathbf W_i$ are designer parameters shaping the closed-loop transfer. At this point it is important to note that the selection of $\mathbf W_o$ and $\mathbf W_i$ can be viewed as an "art" since they completely affect the result in an indirect way. Indeed, engineers are often required to modify the weights and re-optimise, observe the result, and keeping iterating until the expected solution is reached. Still, as we will see in the considered example, some intuitions can be felt when practicing a bit.  In the following code, and keeping in mind the tracking objective we construct such a generalised plant $\mathbf T(\Htranr,\mathbf{\tilde K})=\mathcal F_l(\mathbf P,\mathbf{\tilde K})$. This is done by first defining the $\mathbf P$ (\code{P}) operator as follows.
\begin{lstlisting}[firstline=113,lastline=125,caption={Example file \codeScript{start2ndOrder.m}: $\mathbf P$ (\code{P}) matrix transfer construction.}]
% Construction of the generalised plant embedding the performances
alpha        = 1e3;
wc           = 1;
Wu           = tf([1/wc 1],[1/(wc*alpha) 1]); % weight the control signal
We           = 10/tf([1 0],[1/1e0 1]);        % weight the tracking error signal
systemnames  = 'Hr We Wu';       % declare the dynamical systems
inputvar     = '[r; u]';         % declare the input signals sorted as [w u]
outputvar    = '[Wu; We; r-Hr]'; % declare the output signals sorted as [z y]
input_to_Hr  = '[u]';            % input of the system Hr
input_to_Wu  = '[u]';            % input of the control weigh Wu
input_to_We  = '[r-Hr]';         % input of the tracking error weight We
cleanupsysic = 'yes';            % remove from workspace sysic variables
P            = sysic;            % create the sysetm interconnection
\end{lstlisting}

Following the above code, we define and interconnect two functions $\mathbf W_u$ (\code{Wu}) and  $\mathbf W_e$ (\code{We}) as weights on the control signal $\u$ and error signal $\mathbf e$ respectively. These weights lead to the new fictive outputs  $\mathbf z_u$ and  $\mathbf z_e$ and the resulting generalised model $\mathbf P$ now embeds the following input-output transfers
\begin{equation}
\vectorthree{\mathbf z_u}{\mathbf z_e}{\mathbf r-\y}
=  \left[\begin{array}{cc}
0 & \mathbf W_u \\
\mathbf W_e & -\mathbf W_e\Htranr\\
1 & -\Htranr
\end{array} \right]  \vectortwo{\mathbf r}{\u}
= \mathbf P \vectortwo{\mathbf r}{\u},
\label{eq:P}
\end{equation}
where $\mathbf P \in \Cplx^{3\times 2}$ is a complex-valued matrix function completely defined by the model $\Htranr$ and the weights defined as
$$
\mathbf W_u = \frac{s+1}{s/1000+1} \text{ and }
\mathbf W_e = 10\frac{s+ 1}{s},
$$
being a high pass filter with cut-off frequency at $2\pi$Hz and an integral-like with cut-off also at $2\pi$Hz, respectively (we will come back later on the reason for using such performances). With reference to \eqref{eq:P}, reader may note that the first two outputs are the performances on the control signal tracking error respectively and that the last output is the measurement. Similarly, the first input is the reference while the second one is the control signal. A more systematic way to represent $\mathbf P$ is then given as
$$
\vectortwo{\mathbf z}{\mathbf e} = \mathbf P \vectortwo{\mathbf w}{\u},
$$
where $\mathbf z$ is the performance output vector gathering the variables to control, or more specifically to minimise (here $\mathbf z_u$ and $\mathbf z_e$) and $\mathbf w$ the exogenous signals vector gathering references, disturbances (here only the reference $\mathbf r$). Now $\mathbf P$ has been described, let us define \code{Ktilde}, the controller $\mathbf{\tilde K}$ structure we want to optimise. Here, without entering into technical considerations, we chose a \siso (we measure $\mathbf e=\mathbf r-\y$ and control $\u$) controller of dimension $n_c=2$. In addition we select a controller with no direct feedthrough, \eg a full $\E$ matrix. The following \matlab code stands.
\begin{lstlisting}[firstline=127,lastline=133,caption={Example file \codeScript{start2ndOrder.m}: definition of the controller $\mathbf{\tilde K}$ (\code{Ktilde}) structure. This script defines in some sense the $\mathcal K$ function space, giving the admissible $\mathbf{\tilde K}$ set.}]
% Controller structure definition
ncon          = 1; % number of control variables
nmeas         = 1; % number of measured variables
nc            = 2; % number of internal variables
Ktile         = ltiblock.ss('Ktilde',nc,ncon,nmeas);
Ktile.d.Free  = zeros(ncon,nmeas); 
Ktile.d.Value = zeros(ncon,nmeas);
\end{lstlisting}

Previously, in \eqref{eq:hinfGene}, the space $\mathcal K$ of admissible controllers was introduced. Such a space is somehow defined with the above code by considering the space of \siso controllers of dimension two with no direct feed-through. More specifically, one seeks a controller $\mathbf K$ embedding a state-space model as
$$
\pare{\matrixtwo{1}{0}{0}{1},\matrixtwo{a_1}{a_2}{a_3}{a_4},\vectortwo{b_1}{b_2},\vectortwoT{c_1}{c_2},0},
$$
where all coefficients $a_i$, $b_i$ and $c_i$ are real and considered as design variables. Up to now, one has $\mathbf P$, being a known transfer matrix solely defined by the model $\Htranr$ and the weighting functions ($\mathbf W_i=1$ and $\mathbf W_i=\textbf{blkdiag}(\mathbf W_u,\mathbf W_e)$), and $\mathbf{\tilde K}$, a structured controller with gains to be optimised. Then we have all the ingredients to set up our (once again modified) $\Hinf$ control problem as follows.
\begin{lstlisting}[firstline=135,lastline=143,caption={Example file \codeScript{start2ndOrder.m}: construction of the extended generalised plant $\mathbf T$ (both \code{T} and \code{Text}) and optimal values of $\mathbf K$ (\code{K}).}]
% Model-based Hinf-controller oprimisation (hinfstruct)
T                  = lft(P,Ktile);       % lower LFT
Wk                 = 1e-9*tf(1,1);       % almost unconstrained gain
Text               = append(T,Wk*Ktile); % enforce 'K' stability
option             = hinfstructOptions('Display','iter', ...
                                       'RandomStart',0, ...
                                       'MaxIter',500);
[CLopt,gamma,info] = hinfstruct(Text,option);
K                  = ss(CLopt.Blocks.Ktilde);
\end{lstlisting}

More in details, the above code now considers an extended version of the $\Hinf$ problem, slightly different to \eqref{eq:hinfGene}, formulated as follows.
\begin{equation}
\mathbf K := \arg \min_{
\small
\begin{array}{c}
\mathbf{\tilde K}\in\Hinf \\
\mathbf{\tilde K}\in \mathcal K
\end{array}
\normalsize
}
\begin{array}{c}
\norm{\mathbf T(\Htranr,\mathbf{\tilde K})}_{\Hinf}\\
\norm{\mathbf{\tilde K} \mathbf W_k}_{\Hinf}
\end{array}
\label{eq:hinfGene2}
\end{equation}
where $\mathbf W_k$ is an additional weighting function (here a simple gain) applied directly on the controller $\tilde{\mathbf K}$ to enforce the transfer $\tilde{\mathbf K}\mathbf W_k$ to be stable. In the case described by \eqref{eq:hinfGene2}, the two performance channels ${\mathbf T(\Htranr,\mathbf{\tilde K})}$ and ${\mathbf{\tilde K} \mathbf W_k}$ are appended, leading to the \code{Text} variable, gathering the two objectives. Problem \eqref{eq:hinfGene2} is (NP-)hard to solve (so was also \eqref{eq:hinfGene}) and has been the subject of many research contributions. However, thanks to the developments of the \code{hinfstrut} routine embedded in the \matlab software based on the seminal contribution \cite{Apkarian:2006}, a numerically robust solution can be obtained in a resonable time. The above code first computes the extended generalised variable \code{Text} and then calls the \code{hinfstruct} routine to obtain \code{K}, the optimal controller $\mathbf K$, solving problem \eqref{eq:hinfGene2}.

Then, such code leads to the following informations (note that values may differ from a version and computer to an other).
\begin{lstlisting}
Iter 1: Objective = 347.3, Progress = 100%
Iter 2: Objective = 327.5, Progress = 5.7%
...
Iter 71: Objective = 14.94, Progress = 8.3e-05%
Final: Peak gain = 14.9, Iterations = 71
       Some closed-loop poles are marginally stable (decay rate near 1e-07)
Warning: Gain goal: Feedback configuration has fixed
integrators that cannot be stabilized with available tuning
parameters. Make sure these are modeling artifacts rather
than physical instabilities.
\end{lstlisting}

The last warning is not an issue in our case. Indeed, it states that some instabilities cannot be controlled. In our case, this is caused by the $\mathbf W_e$ weight function exhibiting an integral action and thus $\mathbf P$ has an eigenvalue in zero being uncontrollable.

\begin{figure}[h]
    \centering
    \scalebox{.8}{\tikzstyle{block} = [draw, thick,fill=bleuONERA!20, rectangle, minimum height=3em, minimum width=6em,rounded corners]
\tikzstyle{block2} = [thick,rectangle, minimum height=3em, minimum width=6em,rounded corners]
\tikzstyle{sum} = [draw, thick,fill=bleuONERA!20, circle, node distance=1cm]
\tikzstyle{input} = [coordinate]
\tikzstyle{output} = [coordinate]
\tikzstyle{pinstyle} = [pin edge={to-,thick,black}]
\tikzstyle{connector} = [->,thick]

% The block diagram code is probably more verbose than necessary
\begin{tikzpicture}[auto, node distance=2cm,>=latex']
    % We start by placing the blocks
    \node [input, name=input] {};
    \node [sum, right of=input] (sum) {};
    \node [block, right of=sum, node distance=3cm] (controller) {$\mathbf K(s)$};
    %\node [pinstyle, above of=controller, node distance=1.2cm] (samplingK) {$f_2$};
    %\draw [connector] (samplingK) -- node[name=hK] {} (controller.90);
    %\node [block, right of=controller, node distance=4cm] (pwm) {\textbf{Modulation}};
    %\node [pinstyle, above of=pwm, node distance=1.2cm] (samplingPWM) {$f_1=N f_2$};
    %\draw [connector] (samplingPWM) -- node[name=hPW] {} (pwm.90);
    \node [block, right of=controller, node distance=4cm] (system) {$\Gtran(s)$};
    %\node [pinstyle, above of=system, node distance=1.2cm] (samplingSystem) {$f_1=N f_2$};
    %\draw [connector] (samplingSystem) -- node[name=hPW] {} (system.90);
    \draw [connector] (controller) -- node[name=u] {${\u(t)}$} (system);
    \node [output, right of=system, node distance=3cm] (output) {};

    % Once the nodes are placed, connecting them is easy. 
    \draw [connector] (input) -- node {$\mathbf r(t)$} (sum);
    \draw [connector] (sum) -- node {$\mathbf e(t)$} (controller);
    %\draw [connector] (pwm) -- node {\red{$\mathbf u(t_{k/N})$}} (system);
    \draw [connector] (system) -- node [name=y] {${\y(t)}$} (output);
    \draw [connector] (output)+(-1cm,0) -- ++(-1cm,-2cm) -| node [near start] {} (sum.south);
\end{tikzpicture}}
    \caption{Closed-loop scheme of $\mathbf T_{\mathbf r\y}(\Gtran,\mathbf K)$, being the interconnection of $\mathbf K$ with model $\Gtran$, involved in the validation step.}
    \label{fig:closedLoopContinuousG}
\end{figure}
\begin{figure}[H]
    \centering
    \includegraphics[width=.6\columnwidth]{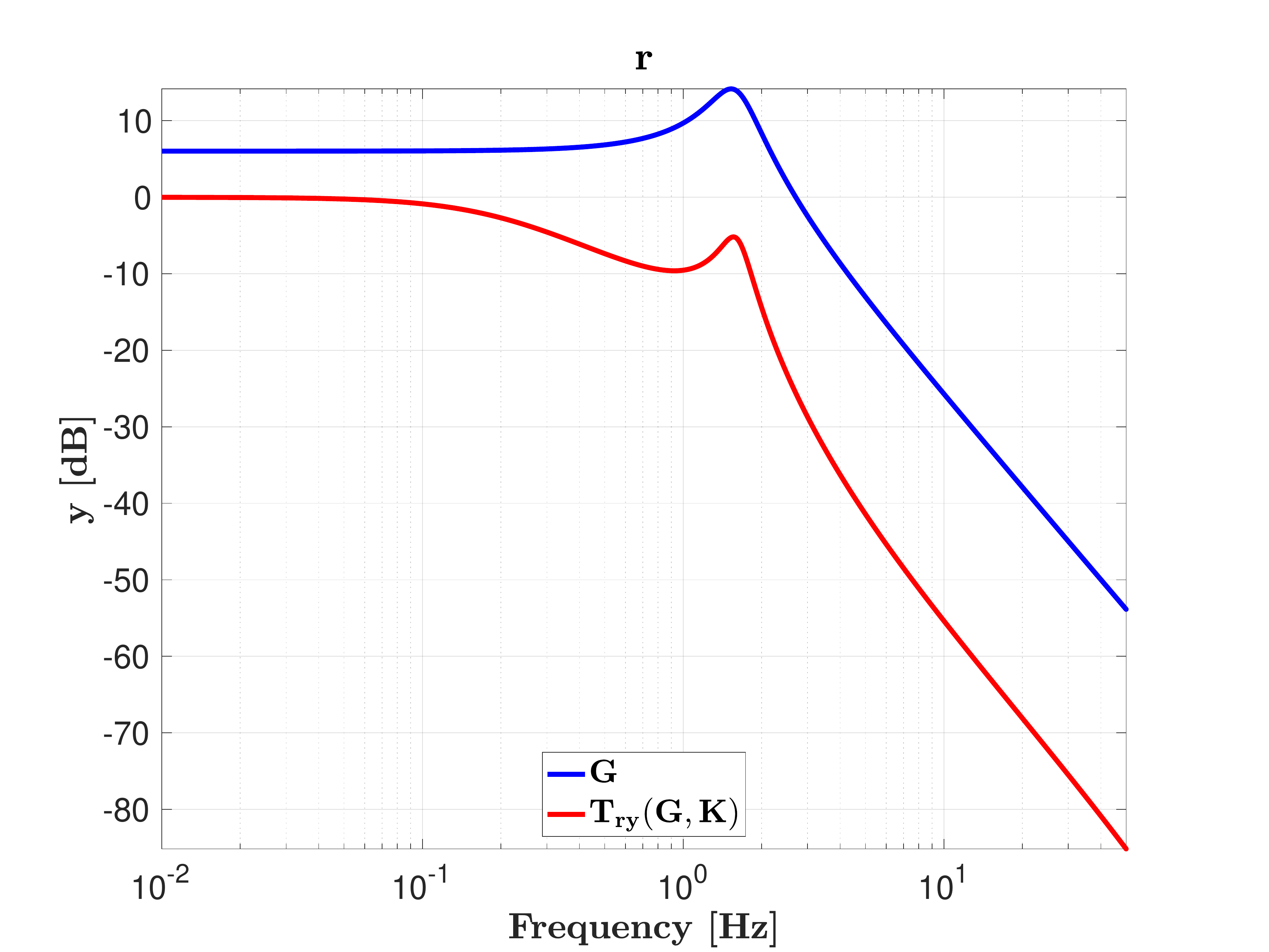}
    \caption{Bode gain of the original model $\Gtran$ (blue) and closed-loop $\mathbf T_{\mathbf r\y}(\Gtran,\mathbf K)$ (red). Note the static gain and peak damping.}
    \label{fig:bodeCL}
\end{figure}
\begin{figure}[H]
    \centering
    \includegraphics[width=.6\columnwidth]{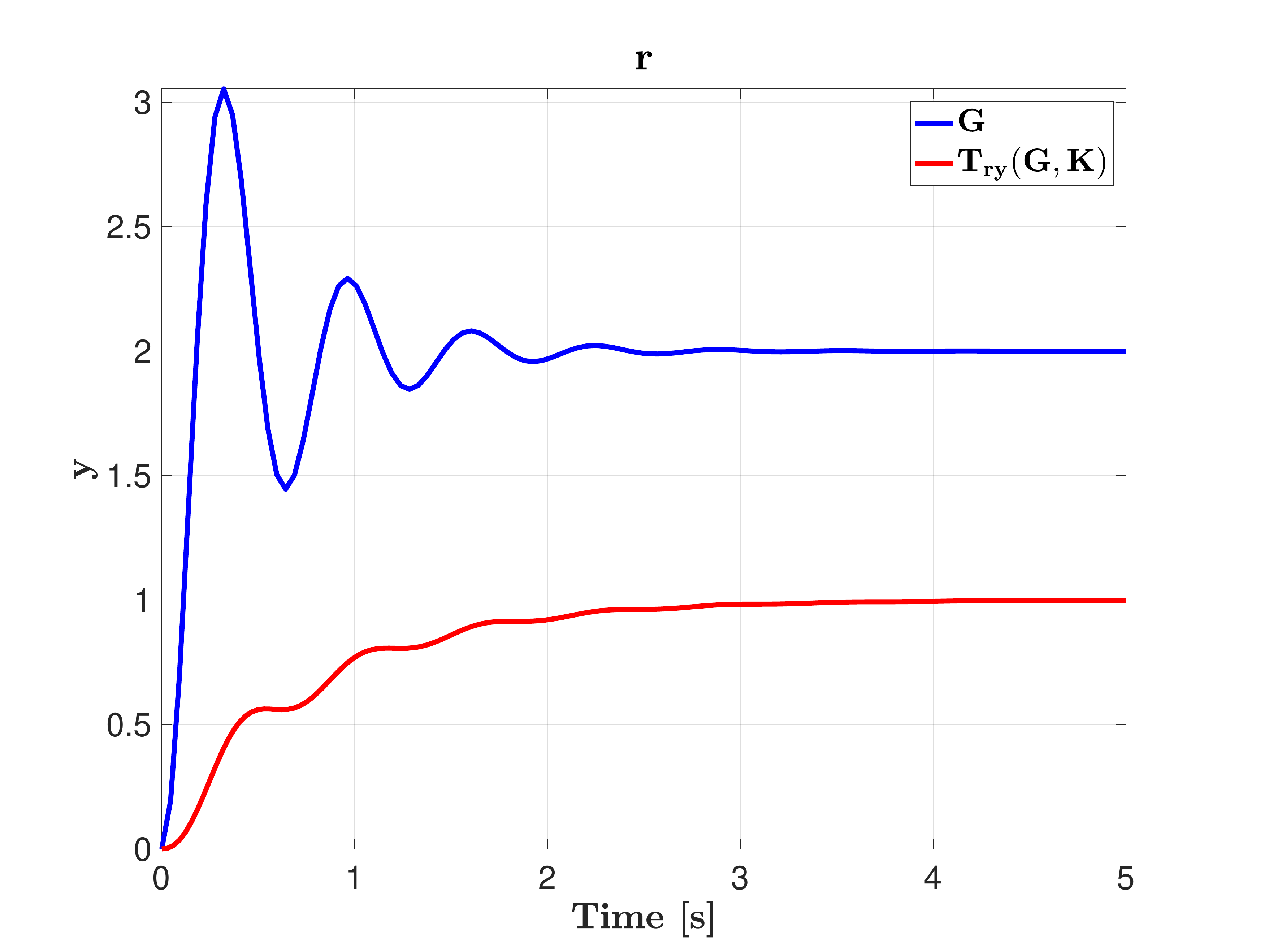}
    \caption{Step response of the original model $\Gtran$ (blue) and closed-loop $\mathbf T_{\mathbf r\y}(\Gtran,\mathbf K)$ (red). Note the static gain and peak damping.}
    \label{fig:stepCL}
\end{figure}

Now that an optimal controller $\mathbf K$ (denoted \code{K}) has been obtained, we can analyse the resulting closed-loop. More specifically, the following code constructs the closed-loop as shown on Figure \ref{fig:closedLoopContinuousG}, involving the original model $\Gtran$. The associated Bode gain and step responses are shown on Figures \ref{fig:bodeCL} and \ref{fig:stepCL}.
\begin{lstlisting}[firstline=145,lastline=153,caption={Example file \codeScript{start2ndOrder.m}: construction of te closed-loop model when original system $\Gtran$ looped with $\mathbf K$.}]
% Closed-loop created with G instead of Hr, looped with K (continuous-time)
systemnames   = 'G K';
inputvar      = '[r]';
outputvar     = '[G; r-G; K]';
input_to_G    = '[K]';
input_to_K    = '[r-G]';
cleanupsysic  = 'yes';
CL            = sysic;
CL            = balreal(CL);
\end{lstlisting}

\begin{figure}[H]
    \centering
    \includegraphics[width=.6\columnwidth]{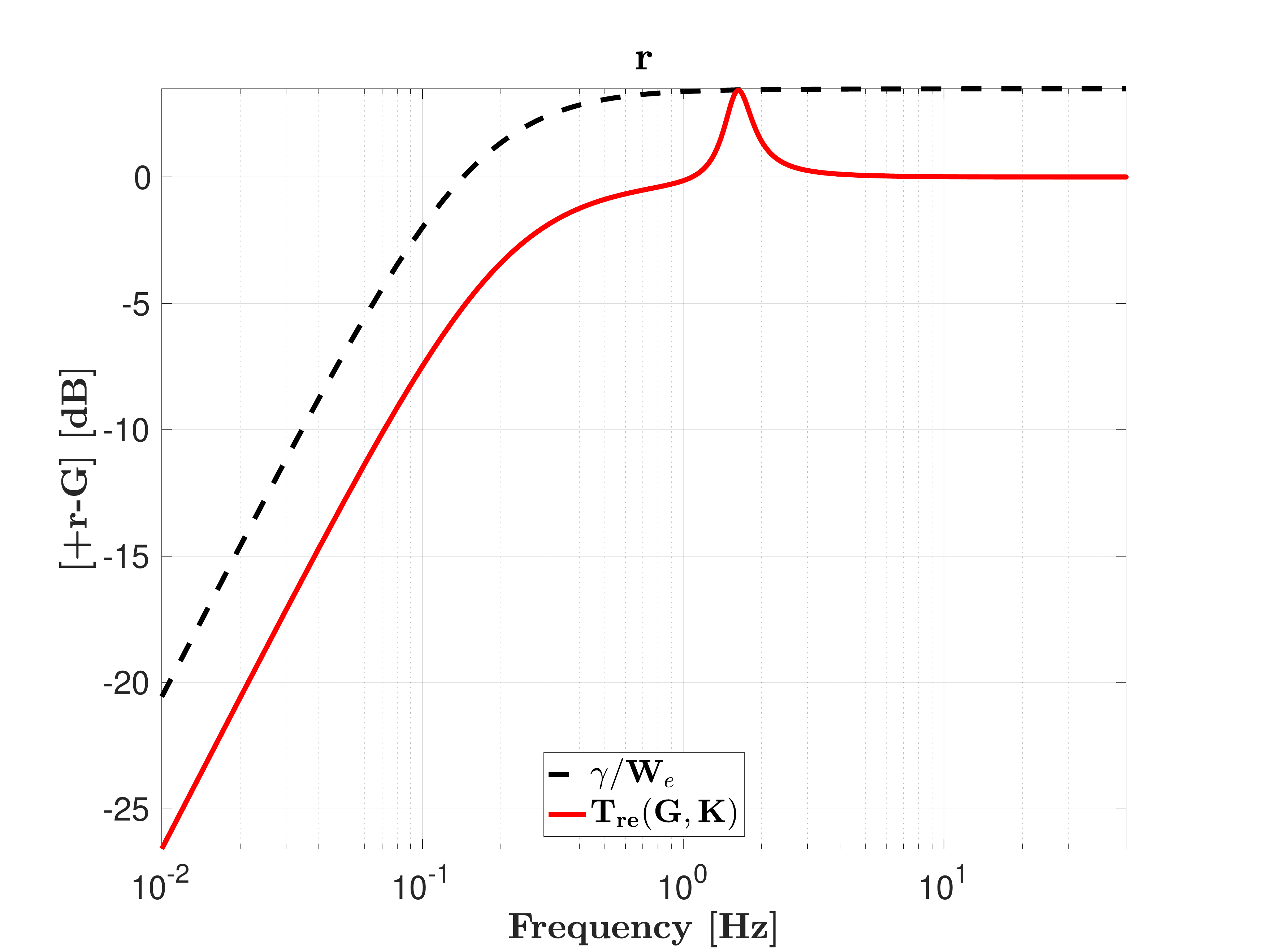}
    \caption{Bode gain of the weighting function on the tracking performance $\gamma/\mathbf W_e$ (black dashed) and closed-loop $\mathbf{T_{rz_e}}(\Gtran,\mathbf K)$ (red).}
    \label{fig:We}
\end{figure}

\begin{figure}[H]
    \centering
    \includegraphics[width=.6\columnwidth]{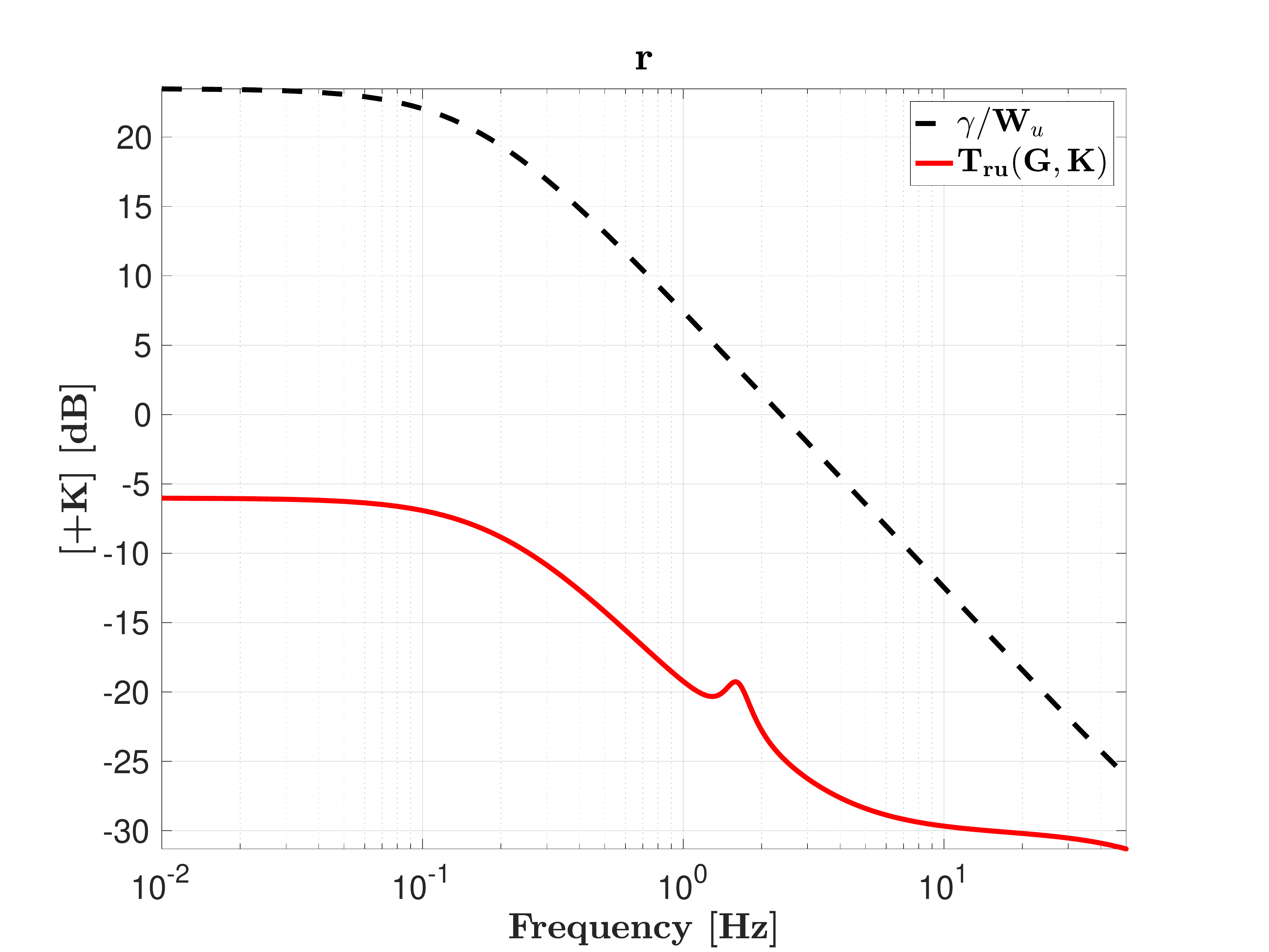}
    \caption{Bode gain of the weighting function on the control performance $\gamma/\mathbf W_u$ (black dashed) and closed-loop $\mathbf{T_{rz_u}}(\Gtran,\mathbf K)$ (red).}
    \label{fig:Wu}
\end{figure}

One observes on  Figure \ref{fig:bodeCL} that the bump present on te open-loop original $\Gtran$ model has been attenuated (\ie damped) and the static gain is now at 0dB, meaning that the output $\y$ should track $\mathbf r$ in steady-state. In addition, Figure  \ref{fig:stepCL} assesses these observations, showing the step response of $\Gtran$ and $\mathbf{T_{ry}}$ obtained using $\Gtran$ looped with $\mathbf K$. One important validation when applying $\Hinf$ control design is the validation of the weighting constraints. This is done on Figures \ref{fig:We} and \ref{fig:Wu}, where the transfer $\mathbf{T_{re}}$ and $\mathbf{T_{ru}}$ are plotted, and compared to the weighting functions $\gamma/\mathbf W_e$ and $\gamma/\mathbf W_u$ respectively (where $\gamma$ denotes  the $\Hinf$ norm obtained when solving \eqref{eq:hinfGene2}).

First, one should note that both transfers are upper bounded by the corresponding weighting functions (black dash dotted), assessing that the considered frequency templates $\mathbf{T(\Htranr,K)}$ (indeed $\mathbf{T(\Gtran,K)}$) are satisfied, \ie $\norm{\mathbf{T(\Gtran,K)}}_{\Hinf} \leq \gamma/\mathbf W_x$ where $x=\{u,e\}$. The second objective being $\norm{\mathbf K}_{\Hinf}\leq \gamma \mathbf W_k$ is also (largely since $W_k=10^{-9}$) satisfied. Remember that such a objective was used only to ensure that the obtained controller $\mathbf K\in\Hinf$, \ie is stable.

At this point, let us just make a quick remark on the section of $\mathbf W_u$ and $\mathbf W_e$. The first one affects the control signal $\u$ and is thus selected so that $1/\mathbf W_u$ rolls-off in high frequencies to avoid noise amplification. Similarly, the $\mathbf W_e$ function, acting on the error signal $\mathbf r-\y$, is standardly selected so that in low frequency $1/\mathbf W_e$ has a low (zero) gain to ensure no steady-state error up to a certain cut-off frequency (being the time response) and constant gain at infinity to monitor the margin performances (as shown in the next part).

\newpage
\subsection{A glimpse of margin}

Entering in a complete margin analysis is not the objective of this simple report (interested reader should refer \eg to \cite{ZhouBook:1997,MagniLFR:2006}). Still, to give a grasp of the concept, margins are generally computed on the following transfer (availability and/or complexity usually make the decision),
$$
\mathbf L(s) = \Gtran(s) \mathbf K(s) \text{  or  } \mathbf L(s) = \Htran(s) \mathbf K(s) \text{  or  } \mathbf L(s) = \Htranr(s) \mathbf K(s)
$$
which represents the loop interconnection without closing the loop. This transfer is actually very important to monitor in practice. Actually, a whole class of design method, called loop-shaping, are rooted on this transfer and aim at shaping it through adequate filtering. Among interesting margin, the modulus margin, denoted \code{ModMargin}, is defined as
$$
\textbf{MM} = \dfrac{1}{\norm{1-\mathbf{T_{ry}}}_{\Hinf}} ,
$$
is a unifying quantity of the gain and phase margins and represents the maximal gain of the so-called sensitivity function, being the transfer from the reference to the error (and thus connected to the weighting function $\mathbf W_e$). These last components can be obtained through the \code{allmargin} routine embedded in \matlab and recalled hereafter.
\begin{lstlisting}[firstline=168,lastline=172,caption={Example file \codeScript{start2ndOrder.m}: some margins.}]
% A glimpse of margin (continuous-time)
L             = G*K;
[reNyq,imNyq] = nyquist(L);
allmargin(L)
ModMargin = 1/norm(CL(2,1),inf);
\end{lstlisting}

Note that complete books are dedicated to the robustness, margin, and related points in the control literature. It is quite hard sorting them, but reader should keep in mind that many numerical tools are embedded in the \matlab software, providing a good starting point.
\begin{lstlisting}
>> allmargin(L)

ans =

  struct with fields:

     GainMargin: 3.5988
    GMFrequency: 11.0321
    PhaseMargin: 90.3163
    PMFrequency: 1.3664
    DelayMargin: 1.1536
    DMFrequency: 1.3664
         Stable: 1
\end{lstlisting}

\begin{figure}[H]
    \centering
    \includegraphics[width=.6\columnwidth]{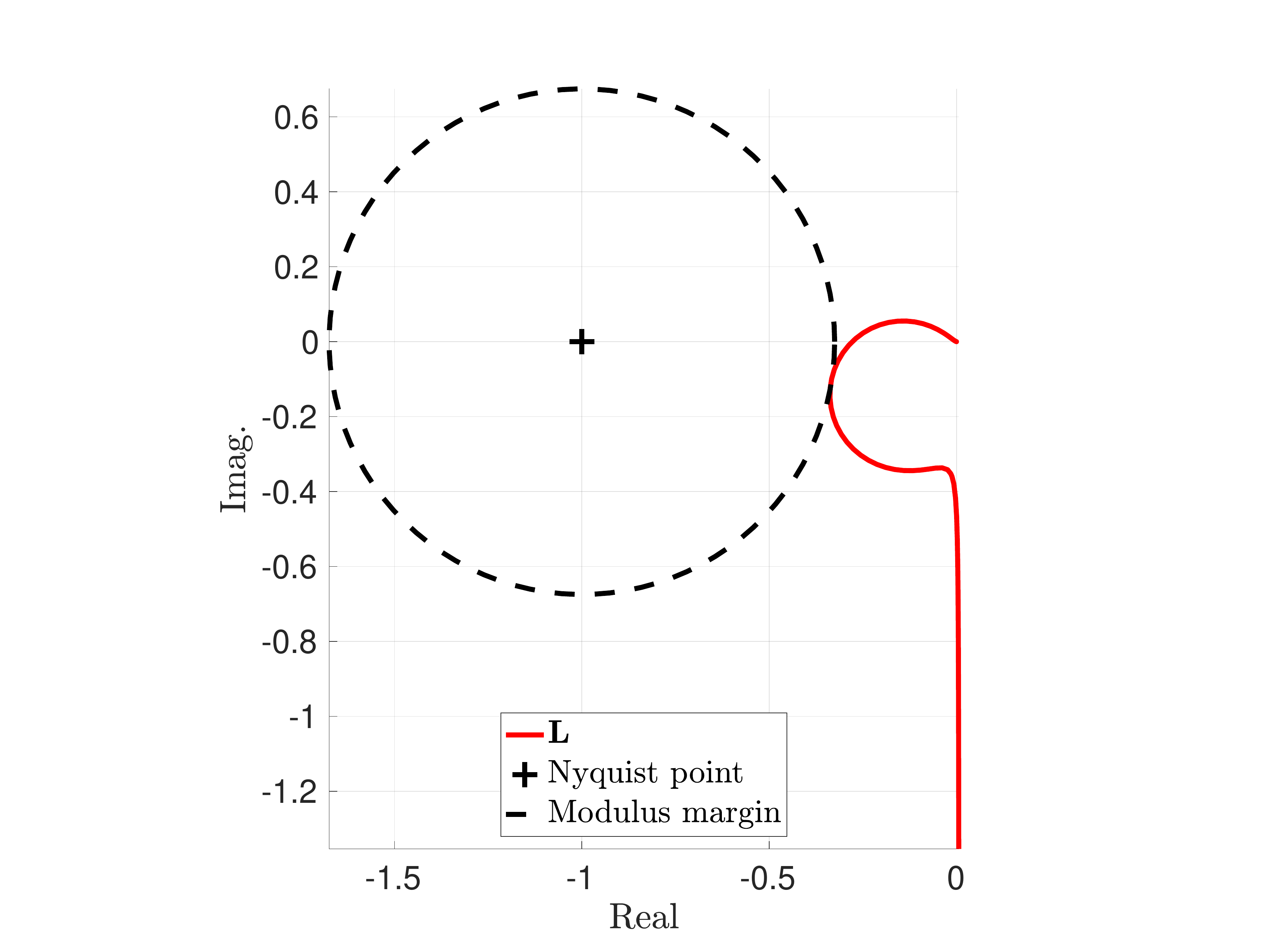}
    \caption{Nyquist plot of the loop transfer $\mathbf L$ (solid red), Nyquist stability point (black $+$) and modulus margin (black dashed circle).}
    \label{fig:nyquist}
\end{figure}

Figure \ref{fig:nyquist} also shows the Nyquist plot of $\mathbf L$, as well as the Nyquist point and the modulus margin illustration. Usually, a modulus margin of $0.5$ is considered as a very good one (note that optimal LQ control for \siso systems ensures this modulus margin).

\subsection{Discrete-time controller and hybrid loop}

Now the optimal continuous-time controller $\mathbf K$ (\code{K}) has been obtained, in view of implementation purpose, it is needed to discretise it to obtain a sampled-system. This step is subject to many research as well and one may refer to \cite{VuilleminDiscreteSub} for some insight. Without being too specific, the standard bilinear Mobius transform (also celebrated as Tustin stransformation) is used. It basically consists in a transformation from the $s$-plane to the $z$-plane using the following formulae
$$
s = \dfrac{2}{T_s}\dfrac{z-1}{z+1},
$$
where $T_e$ is the sampling time. In \matlab this transformation is implemented and can be obtained as follows.
\begin{lstlisting}[firstline=183,lastline=186,caption={Example file \codeScript{start2ndOrder.m}: discretisation step using bilinear transformation.}]
% Controller continuous vs. sampled (discretisation using Matlab)
Ts     = 1e-2;               % sampling time of the controller
K_z    = c2d(K,Ts,'tustin'); % bilinear discretisation method
FR_K_z = freqresp(K_z,W);
\end{lstlisting}

\begin{figure}[H]
    \centering
    \includegraphics[width=.6\columnwidth]{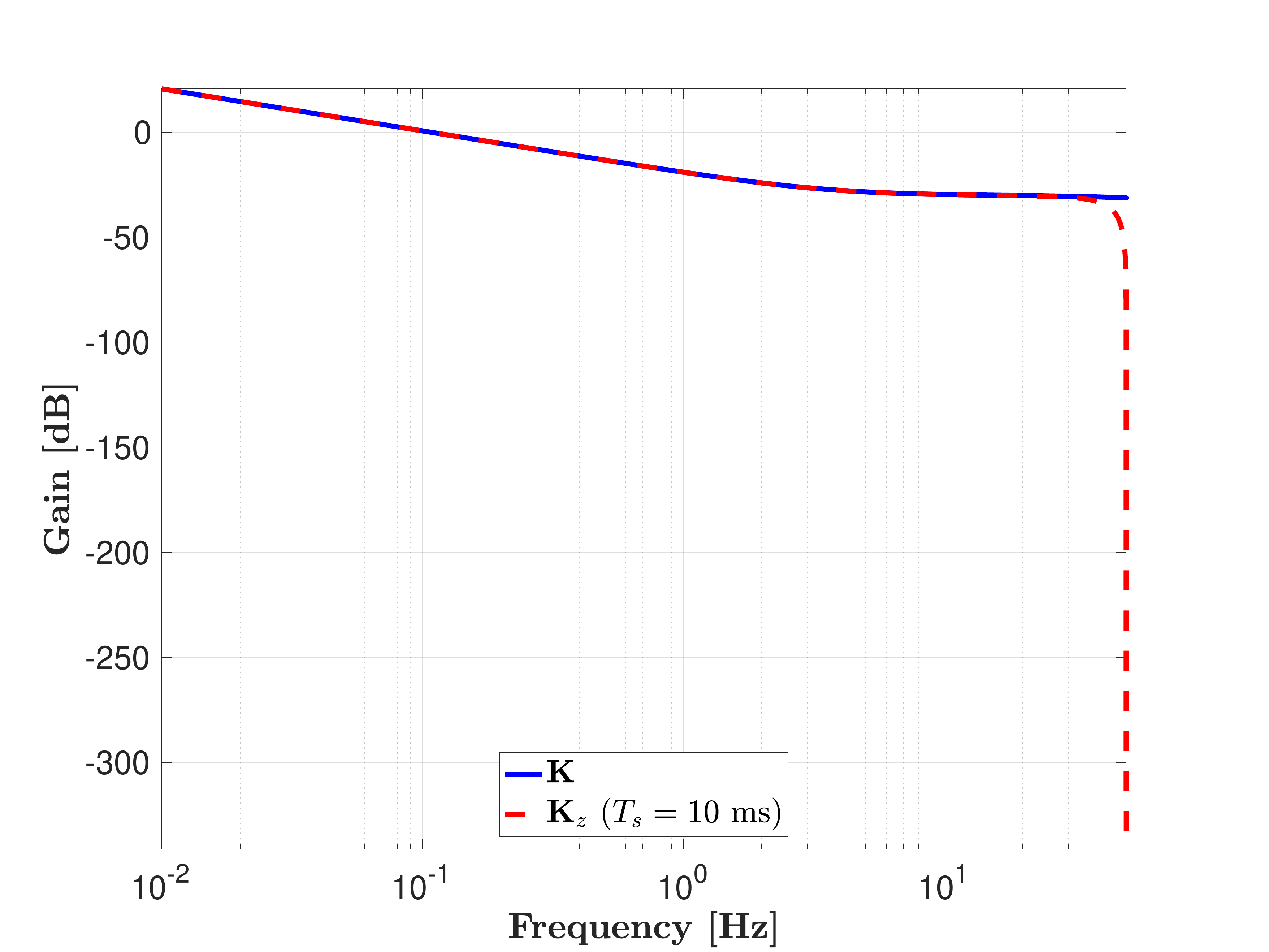}
    \caption{Comparison of the controller Bode response. $\mathbf K$ (\code{K}) in continuous-time (solid blue) and $\mathbf K_z$  (\code{K\_z}) in sampled-time at $T_s=10$ms (red dashed).}
    \label{fig:KzTs10}
\end{figure}

The above code then creates the discrete-time controller $\mathbf K_z$ (denoted \code{K\_z}) obtained with a sampling time $T_s=10$ms defined as follows, and for which frequency response (Bode diagram) is given in Figure \ref{fig:KzTs10}.
\begin{lstlisting}
>> K_z

K_z =

  A =
             x1        x2
   x1         1  0.001432
   x2    0.0013   -0.4662

  B =
              u1
   x1   0.004413
   x2  -0.004758

  C =
           x1      x2
   y1   1.522  -2.466

  D =
            u1
   y1  0.02536

Sample time: 0.01 seconds
Discrete-time state-space model.
\end{lstlisting}

When analysing Figure \ref{fig:KzTs10}, one first note that the continuous-time response of $\mathbf K$ is similar to the $\mathbf K_z$ up to $f_{\text{Nyquist}}=f_s/2=1/(2T_s)=5$Hz. Moreover, as the bilinear transformation maps the vertical axis $\imath\Real$ onto the unit circle, the spectrum is naturally repeated every $f_{\text{Nyquist}}$, resulting in these sharp peaks in the Bode gain responses at this frequency (and its periodic multiplicities).

\begin{figure}[H]
    \centering
    \includegraphics[width=.45\columnwidth]{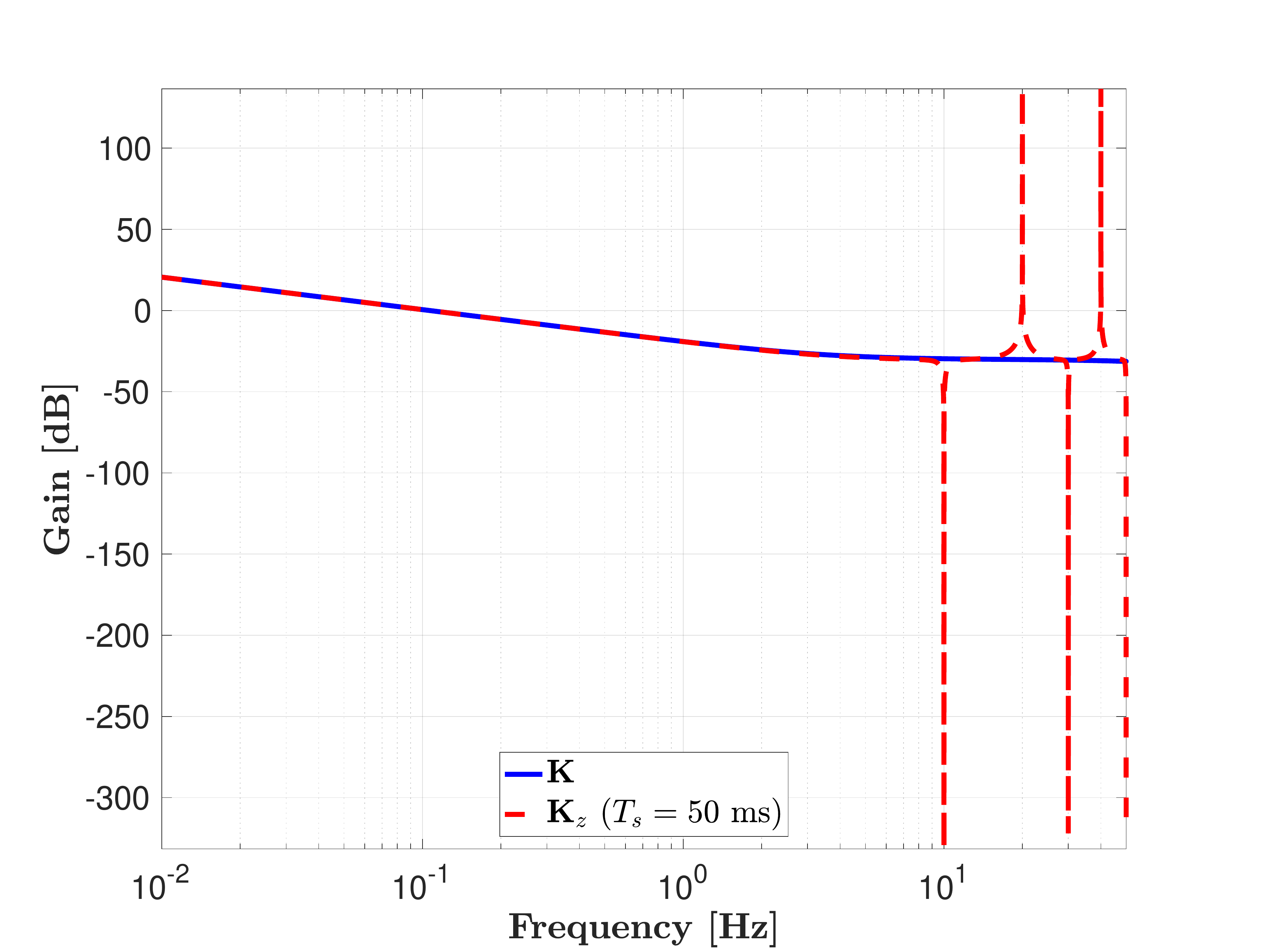}
    \includegraphics[width=.45\columnwidth]{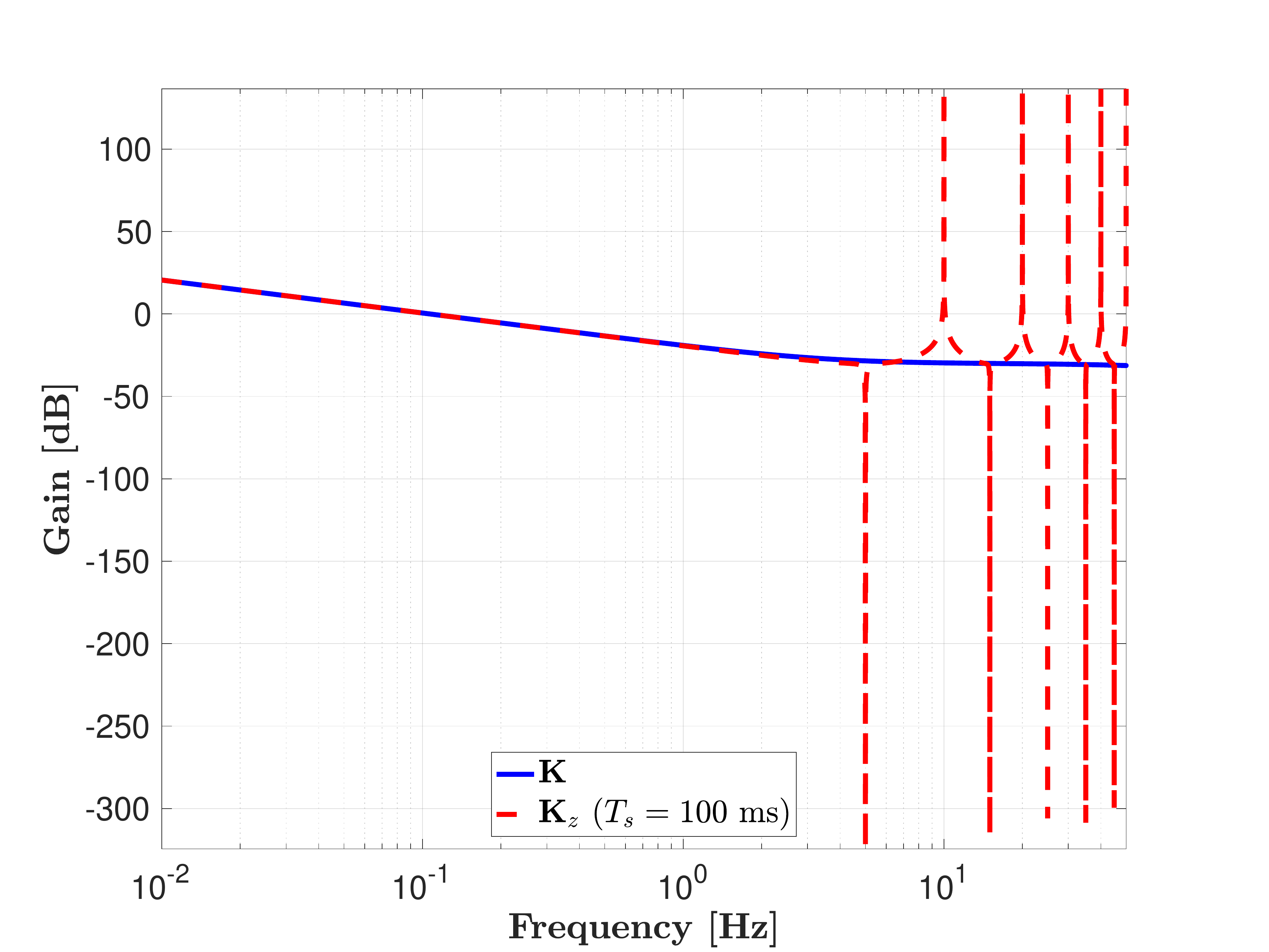}
    \includegraphics[width=.45\columnwidth]{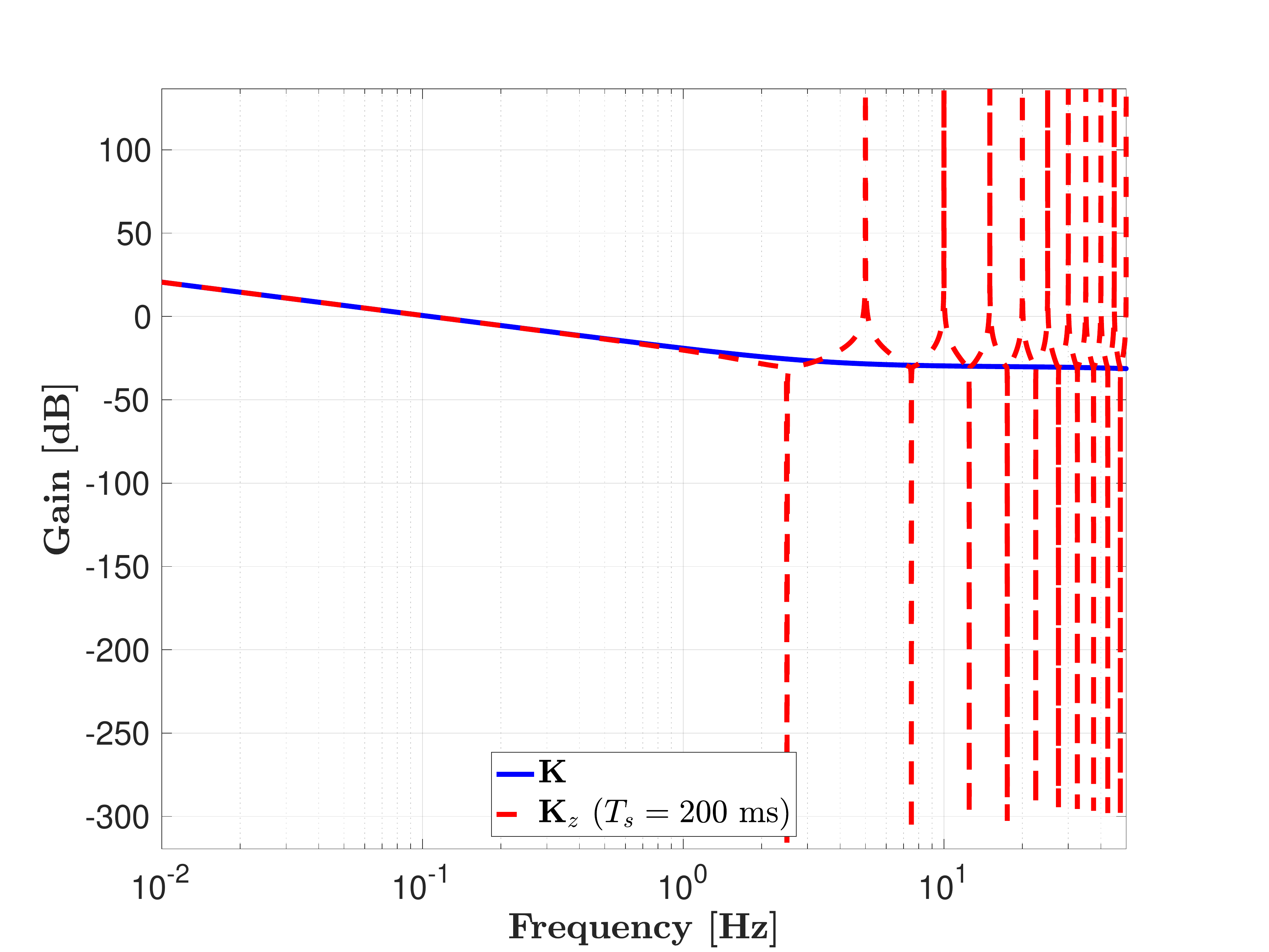}
    \includegraphics[width=.45\columnwidth]{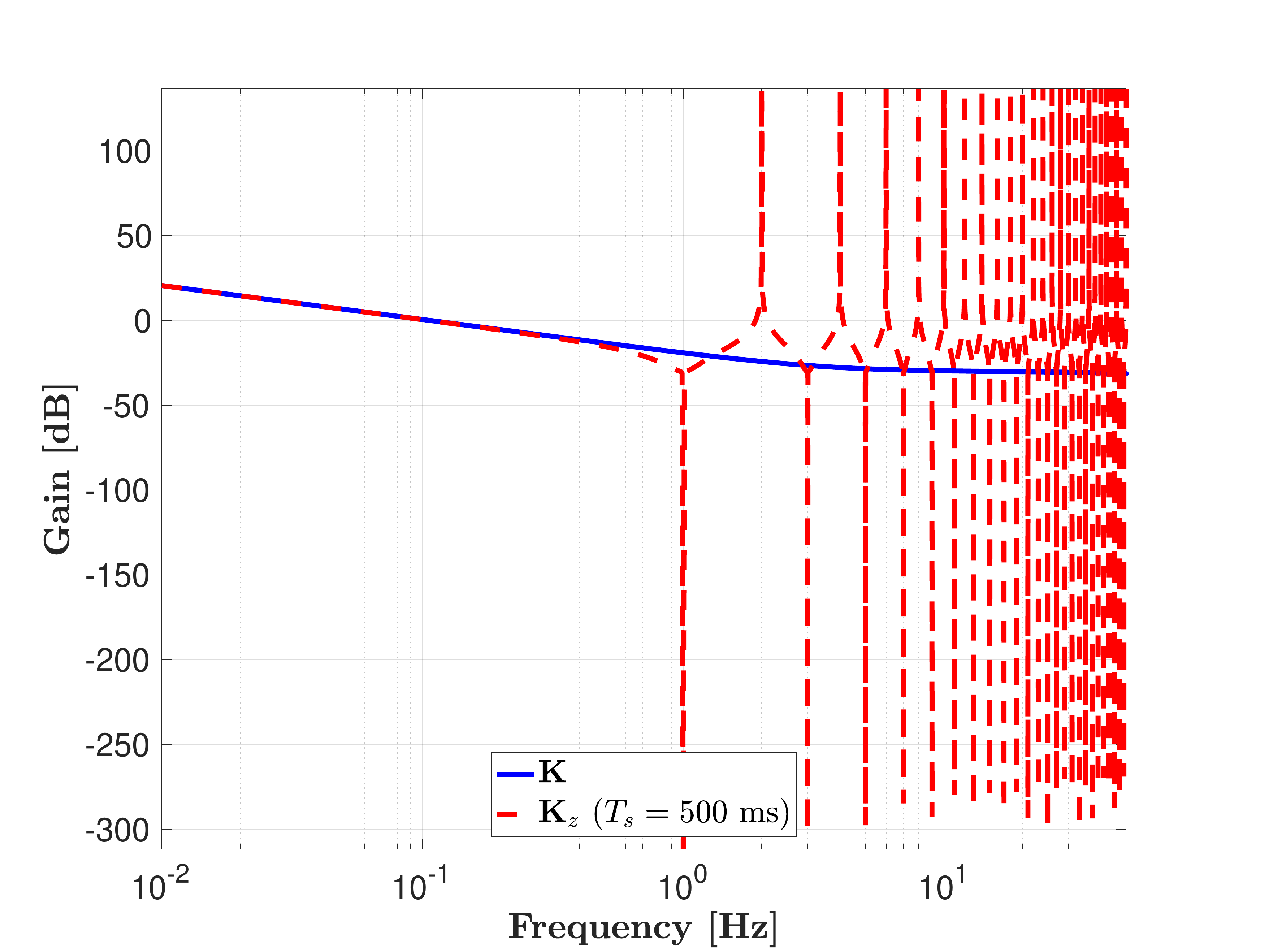}    \caption{Comparison of the controller Bode response. $\mathbf K$ in continuous-time (solid blue) and $\mathbf K_z$ in sampled-time (red dashed). From top left to bottom right $T_s=50$, $T_s=100$, $T_s=200$ and $T_s=500$ms.}
    \label{fig:KzTsVs}
\end{figure}

Obviously, one can reduce the sampling time as much as possible, leading to a better frequency response matching, but with a technical and practical limitation (and cost). A glimpse of the impact of the sampling frequency is shown on Figure \ref{fig:KzTsVs} which highlights the effect of decreasing the sampling frequency. It is clear the slower the discretisation, the less the discrete controller $\mathbf K_z$ matches the (reference) continuous one $\mathbf K$. In addition, the phase plot is also affected, leading to shift and delay in the loop. This point is not reported here, but it may lead to instabilities (as illustrated in the next section). Many more remarks are available in \cite{VuilleminDiscreteSub} and references therein.

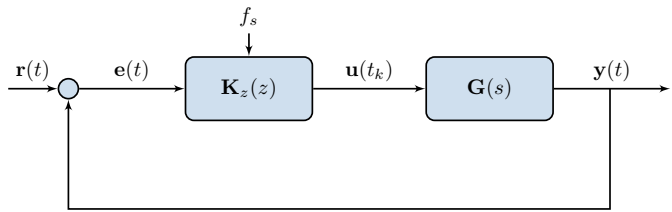
\begin{figure}[H]
    \centering
    \scalebox{.8}{\tikzstyle{block} = [draw, thick,fill=bleuONERA!20, rectangle, minimum height=3em, minimum width=6em,rounded corners]
\tikzstyle{block2} = [thick,rectangle, minimum height=3em, minimum width=6em,rounded corners]
\tikzstyle{sum} = [draw, thick,fill=bleuONERA!20, circle, node distance=1cm]
\tikzstyle{input} = [coordinate]
\tikzstyle{output} = [coordinate]
\tikzstyle{pinstyle} = [pin edge={to-,thick,black}]
\tikzstyle{connector} = [->,thick]

% The block diagram code is probably more verbose than necessary
\begin{tikzpicture}[auto, node distance=2cm,>=latex']
    % We start by placing the blocks
    \node [input, name=input] {};
    \node [sum, right of=input] (sum) {};
    \node [block, right of=sum, node distance=3cm] (controller) {$\mathbf K_z(z)$};
    \node [pinstyle, above of=controller, node distance=1.2cm] (samplingK) {$f_s$};
    \draw [connector] (samplingK) -- node[name=hK] {} (controller.90);
    %\node [block, right of=controller, node distance=4cm] (pwm) {\textbf{Modulation}};
    %\node [pinstyle, above of=pwm, node distance=1.2cm] (samplingPWM) {$f_1=N f_2$};
    %\draw [connector] (samplingPWM) -- node[name=hPW] {} (pwm.90);
    \node [block, right of=controller, node distance=4cm] (system) {$\mathbf G(s)$};
    %\node [pinstyle, above of=system, node distance=1.2cm] (samplingSystem) {$f_1=N f_2$};
    %\draw [connector] (samplingSystem) -- node[name=hPW] {} (system.90);
    \draw [connector] (controller) -- node[name=u] {${\u(t_k)}$} (system);
    \node [output, right of=system, node distance=3cm] (output) {};

    % Once the nodes are placed, connecting them is easy. 
    \draw [connector] (input) -- node {$\mathbf r(t)$} (sum);
    \draw [connector] (sum) -- node {$\mathbf e(t)$} (controller);
    %\draw [connector] (pwm) -- node {\red{$\mathbf u(t_{k/N})$}} (system);
    \draw [connector] (system) -- node [name=y] {${\y(t)}$} (output);
    \draw [connector] (output)+(-1cm,0) -- ++(-1cm,-2cm) -| node [near start] {} (sum.south);
\end{tikzpicture}}
    \caption{Closed-loop scheme of $\mathbf T_{\mathbf r\y}(\Gtran,\mathbf K_z)$, being the interconnection of the sampled $\mathbf K_z$ with model $\Gtran$, involved in the validation step.}
    \label{fig:closedLoopMixed}
\end{figure}

One very complex and interesting question arising at this point is to analyse the impact of such discretisation in the closed-loop performances. With reference to Figure \ref{fig:closedLoopMixed}, the interconnection of the continuous-time model $\Gtran$ with the sampled-time controller $\mathbf K_z$ is an hybrid system blending continuous and discrete variables. The analysis of such an interconnection is much more involved than the study of a purely continuous-time interconnection and dedicated methods, that go way over the scope of this report, are required.
%Its analysis is complicated and as it is not the objective of the report to provide a complex analysis method, let us simply mention that this makes the task very complicated, even for simple simulation purpose.

In the following section, this sampled controller will be interconnected to a modulation box and analysis will be done on this (even more) complex loop. A sketch of solution and the main ideas will be discussed. Theoretical considerations are purposely left aside.

%\newpage
\section{Signal modulation and hybrid closed-loop validation}
\label{sec-modulation}
The points evoked in the previous sections are related to dynamical systems theory with the main objective of constructing a digital sampled control law $\mathbf K_z$ achieving some performances. From now, we consider that such a controller has to be implemented with an additional limitation on the actuator capability. More specifically, here we consider the interconnection of a sampled-time controller with a continuous-time system and the effect of a pulsed width modulation in the actuator.

\subsection{Preliminary words}

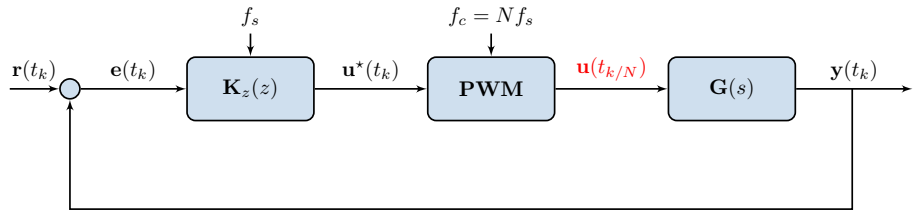
\begin{figure}[h]
    \centering
    \scalebox{.8}{\tikzstyle{block} = [draw, thick,fill=bleuONERA!20, rectangle, minimum height=3em, minimum width=6em,rounded corners]
\tikzstyle{block2} = [thick,rectangle, minimum height=3em, minimum width=6em,rounded corners]
\tikzstyle{sum} = [draw, thick,fill=bleuONERA!20, circle, node distance=1cm]
\tikzstyle{input} = [coordinate]
\tikzstyle{output} = [coordinate]
\tikzstyle{pinstyle} = [pin edge={to-,thick,black}]
\tikzstyle{connector} = [->,thick]

% The block diagram code is probably more verbose than necessary
\begin{tikzpicture}[auto, node distance=2cm,>=latex']
    % We start by placing the blocks
    \node [input, name=input] {};
    \node [sum, right of=input] (sum) {};
    \node [block, right of=sum, node distance=3cm] (controller) {$\mathbf K_z(z)$};
    \node [pinstyle, above of=controller, node distance=1.2cm] (samplingK) {$f_s$};
    \draw [connector] (samplingK) -- node[name=hK] {} (controller.90);
    %\node [block, right of=controller, pin={[pinstyle]above:Disturbances},node distance=3cm] (system) {System};
    \node [block, right of=controller, node distance=4cm] (pwm) {$\textbf{PWM}$};
    \node [pinstyle, above of=pwm, node distance=1.2cm] (samplingPWM) {$f_c=N f_s$};
    \draw [connector] (samplingPWM) -- node[name=hPW] {} (pwm.90);
    \node [block, right of=pwm, node distance=4cm] (system) {$\Gtran(s)$};
    %\node [pinstyle, above of=system, node distance=1.2cm] (samplingSystem) {$f_c=N f_s$};
    %\draw [connector] (samplingSystem) -- node[name=hPW] {} (system.90);
    %\node [block, right of=controller, node distance=3cm] (reference) {$M$};
    % We draw an edge between the controller and system block to 
    % calculate the coordinate u. We need it to place the measurement block. 
    \draw [connector] (controller) -- node[name=u] {${\u^\star(t_k)}$} (pwm);
    \node [output, right of=system, node distance=3cm] (output) {};
    %\node [block, below of=u] (measurements) {Measurements};

    % Once the nodes are placed, connecting them is easy. 
    \draw [connector] (input) -- node {$\mathbf r(t_k)$} (sum);
    \draw [connector] (sum) -- node {$\mathbf e(t_k)$} (controller);
    \draw [connector] (pwm) -- node {\red{$\mathbf u(t_{k/N})$}} (system);
    \draw [connector] (system) -- node [name=y] {${\y(t_k)}$} (output);
    %\draw [->] (y) |-| (measurements);
    %\draw [->] (measurements) -| node[pos=0.99] {$-$} node [near end] {$y_m$} (sum);
    \draw [connector] (output)+(-1cm,0) -- ++(-1cm,-2cm) -| node [near start] {} (sum.south);
\end{tikzpicture}}
    \caption{Closed-loop scheme of $\mathbf T_{\mathbf r\y}(\Gtran,\mathbf K_z+\textbf{PWM})$, being the interconnection of the sampled $\mathbf K_z$ with model $\Gtran$, where control is modulated by the \textbf{PWM}.}
    \label{fig:closedLoopMixedModulated}
\end{figure}

Let us now consider that an actuator, between the system and the controller is no longer able to deliver a continuously value $\u(t)$, or more specifically $\u(t_k)$ (at each sample $T_s$), but a pulsed value with varying duration only, being either $\u_{\textbf{min}}$ or $ \u_{\textbf{max}}$. This case is illustrated on Figure \ref{fig:closedLoopMixedModulated}, where the \textbf{PWM} (Pulsed Width Modulation) block is detailed in what follows. Note that in this section, no \matlab code is provided as this part is still under high investigation from the authors, and description of the code would be quite complicated for this tutorial.

\subsection{The pulsed width modulation case}

The \textbf{PWM} block uses a rectangular impulsion signal taking values between $\u_{\text{min}}$ and $\u_{\text{max}}$ and which length is modulated. This modulation results in variation of the mean $\overline{\u^\star}(t_k)$ of the signal $\u^\star(t_k)$ to convert. If one considers an impulsion with a high frequency $f_c$ and a duty cycle $D\in[0~1]$, the mean value of the resulting signal is given by
\begin{equation}
\begin{array}{rcl}
     \overline{\u^\star}(t_k) &=& \dfrac{1}{T_c} \displaystyle\int_0^{T_c} \u^\star(t_k) dt_k \\
     &=& \dfrac{1}{T_c} \bigg( \displaystyle\int_0^{DT_c} \u_{\text{max}} dt_k + \displaystyle\int_{DT_c}^{T_c} \u_{\text{min}} dt_k \bigg) \\
     &=& D \u_{\text{max}} + (1-D)\u_{\text{min}} \\
     &=& D \u_{\text{max}} \text{ (for $\u_{\text{min}}=0$)}.
\end{array}
\label{eq:pwm}
\end{equation}

Obviously, the \textbf{PWM} should be $N\in\mathbb N$ times higher than the signal $\u^\star(t_k)$ to be modulated. In practical applications, a simple way to generate the \textbf{PWM} is to use the intersection method which simply requires a saw-tooth carrier signal denoted $\u_c(t_k)$, with frequency $f_c=f_s/N$ and amplitude from $\u_{\text{min}}=\min \u^\star(t_k)$ to $\u_{\text{max}}=\max \u^\star(t_k)$, that should be compared to the incoming signal $\u^\star(t_k)$. When $\u_c(t_k)>\u^\star(t_k)$, then $\u(t_{k/N})=\u_{\text{max}}$, and $\u(t_{k/N})=\u_{\text{min}}$ otherwise. In our case, the carrier signal has the same period as the control $\u^\star(t_k)$ and the modulated signal is $N=10$ times faster.

\subsection{Time-domain simulations}

On the basis of the above controller $\mathbf K_z$ connected to a \textbf{PWM} block as the one described in the previous section, following Figure \ref{fig:closedLoopMixedModulated}, we are now ready to perform different time-domain simulations to illustrate the efficiency and limitations of the such an interconnection. In the following Figures \ref{fig:odeY} and  \ref{fig:odeU}-\ref{fig:odeUpwm}, the outputs $\y$ and control signals $\u$, obtained using the different closed-loop schemes and different sampling times ($T_s=\{50,100,200,500\}$ms), namely the one on Figure \ref{fig:closedLoopContinuousG}, \ref{fig:closedLoopMixed} and \ref{fig:closedLoopMixedModulated} are presented. In all cases, we use a \textbf{PWM} block that goes $N=10$ times faster. Moreover, the block provides $\u_{\text{min}}=0$ and $\u_{\text{max}}=1$ only. Note at this point that the amplitude of the \textbf{PWM} also plays an important role, but this is clearly out of the linear domain of this report.

\begin{figure}[H]
    \centering
    \includegraphics[width=.45\columnwidth]{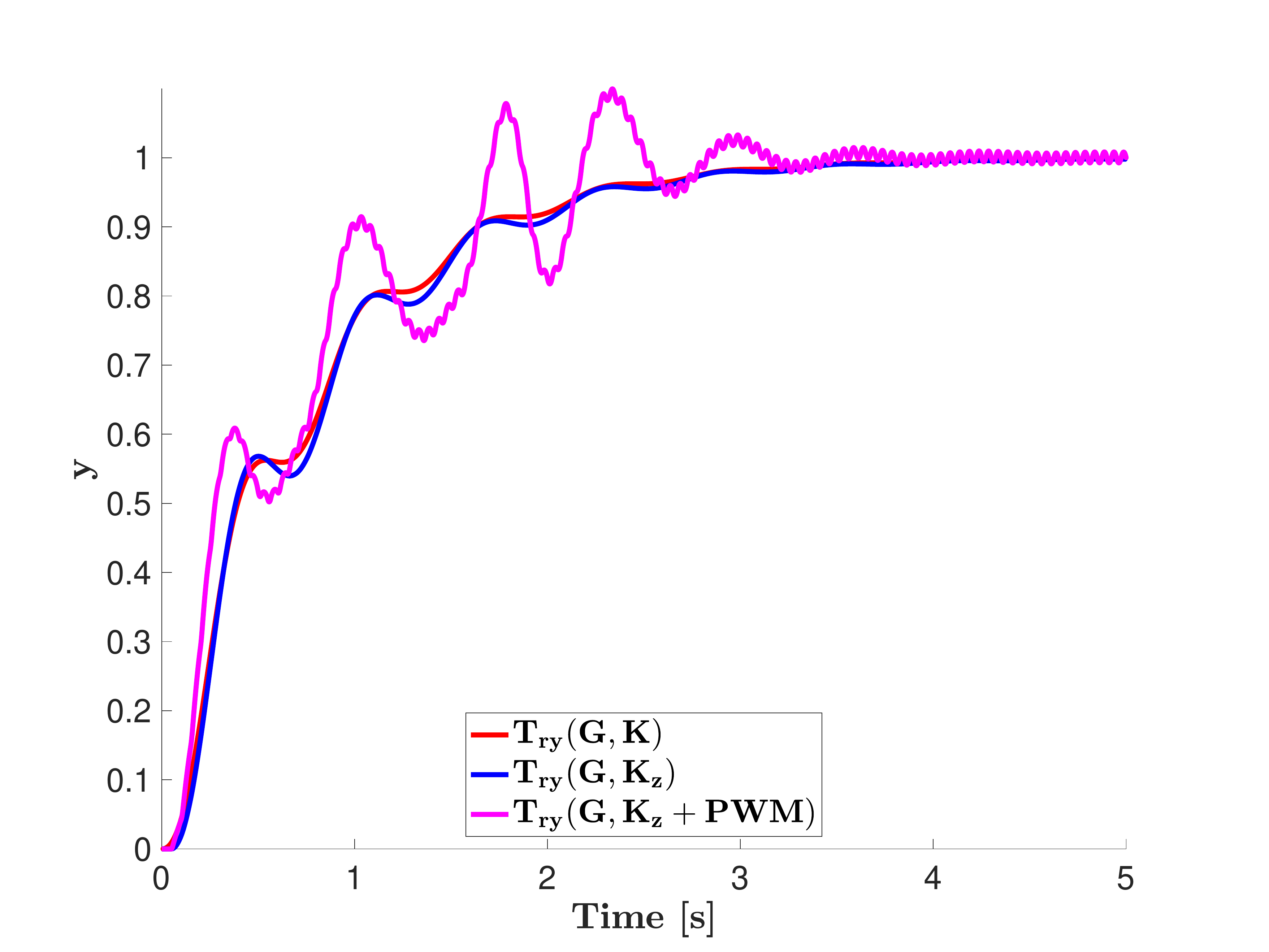}
    \includegraphics[width=.45\columnwidth]{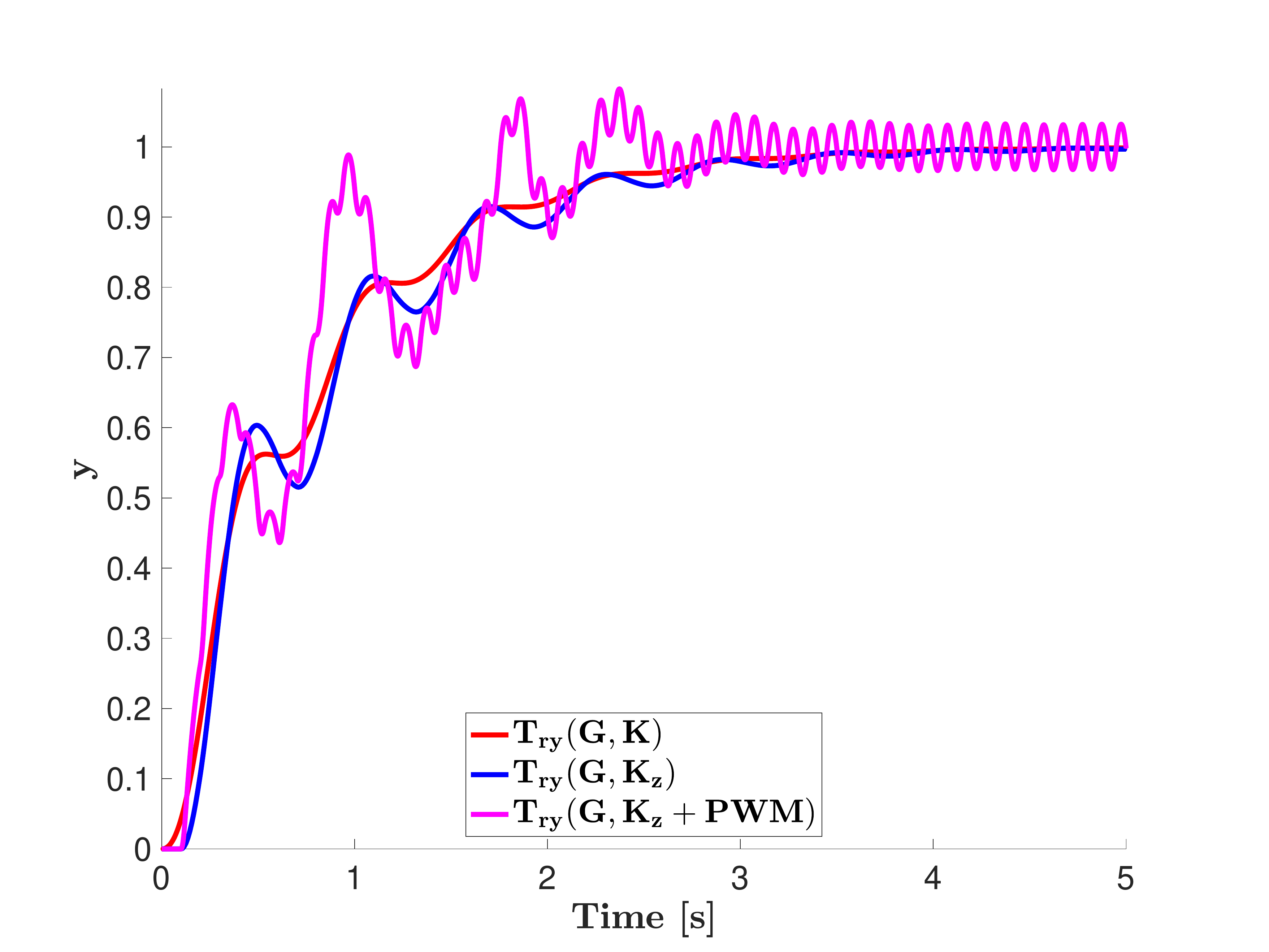} \\
    \includegraphics[width=.45\columnwidth]{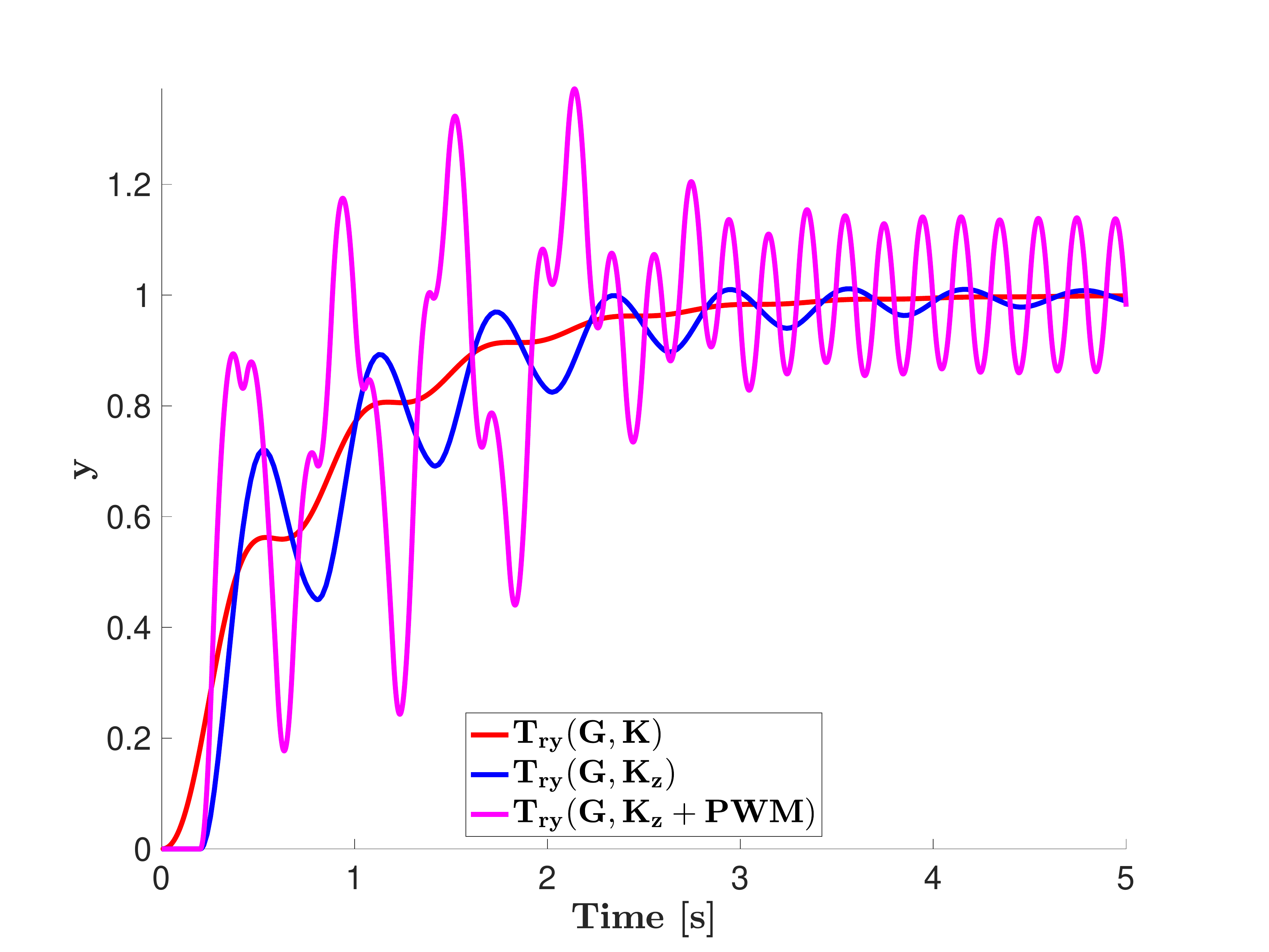}
    \includegraphics[width=.45\columnwidth]{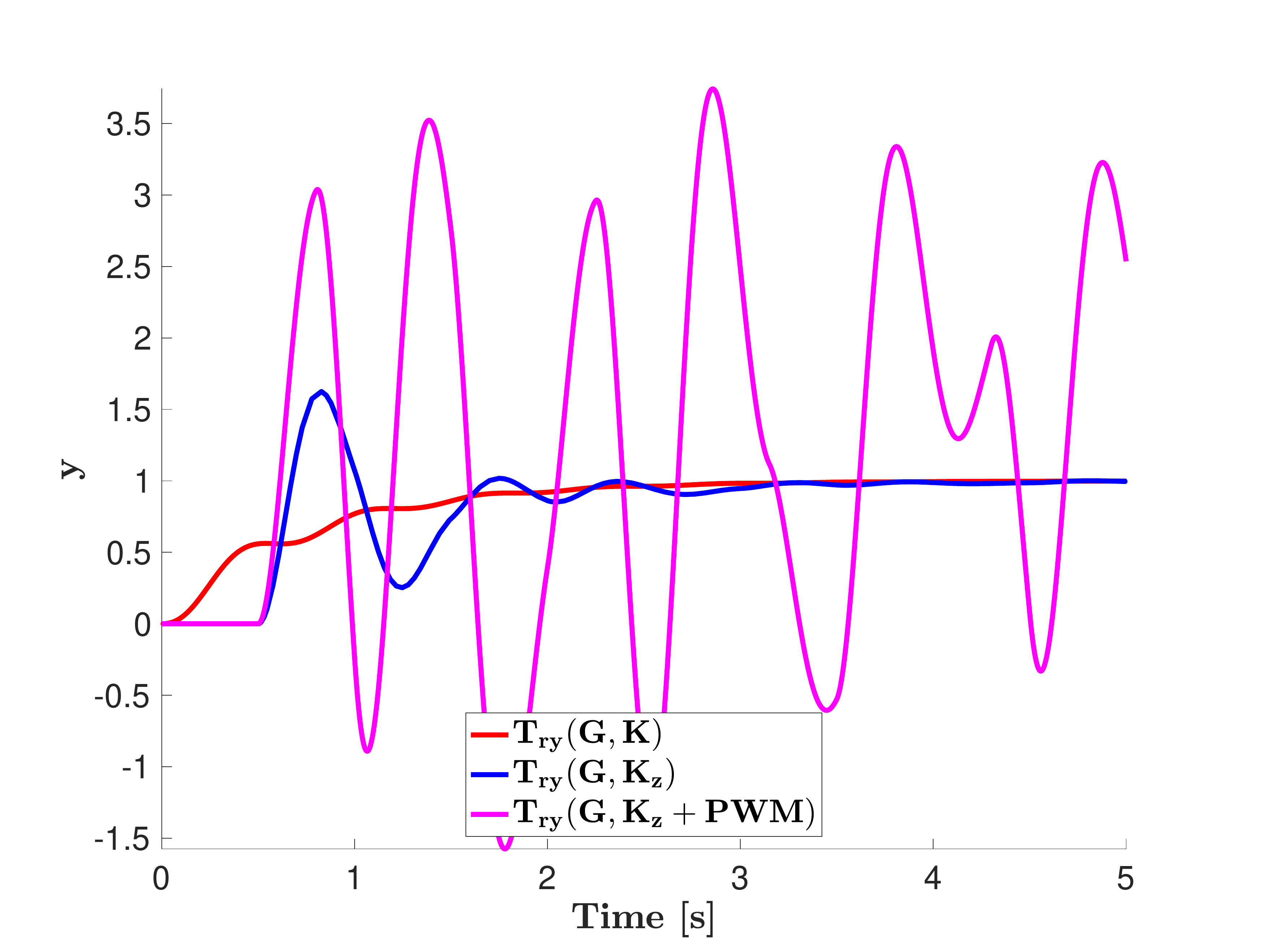}
    \caption{From top left to bottom right: $T_s=50$ms, $T_s=100$, $T_s=200$ and $T_s=500$ms. Comparison of the controller step responses of the different closed-loop scheme. The continuous-time closed-loop $\Gtran$-$\mathbf K$ (solid red), the hybrid continuous-sampled-time $\Gtran$-$\mathbf K_z$ in  (solid blue) and the hybrid continuous-sampled-time and modulated $\Gtran$-$\mathbf K_z$-\textbf{PWM} (solid magenta).}
    \label{fig:odeY}
\end{figure}

Clearly, when sampling time increases, as shown on Figure \ref{fig:odeY}, the hybrid closed-loop is diverging from the continuous one (tuned in the previous section). This is first observed by comparing the red curve (reference) with the blue one (hybrid loop), where a loss of tracking performance is visible. This observation is even more visible when the control signal is modulated using the \textbf{PWM} block. In the last case ($T_s=500$ms), the closed-loop even becomes unstable!

\begin{figure}[H]
    \centering
    \includegraphics[width=.45\columnwidth]{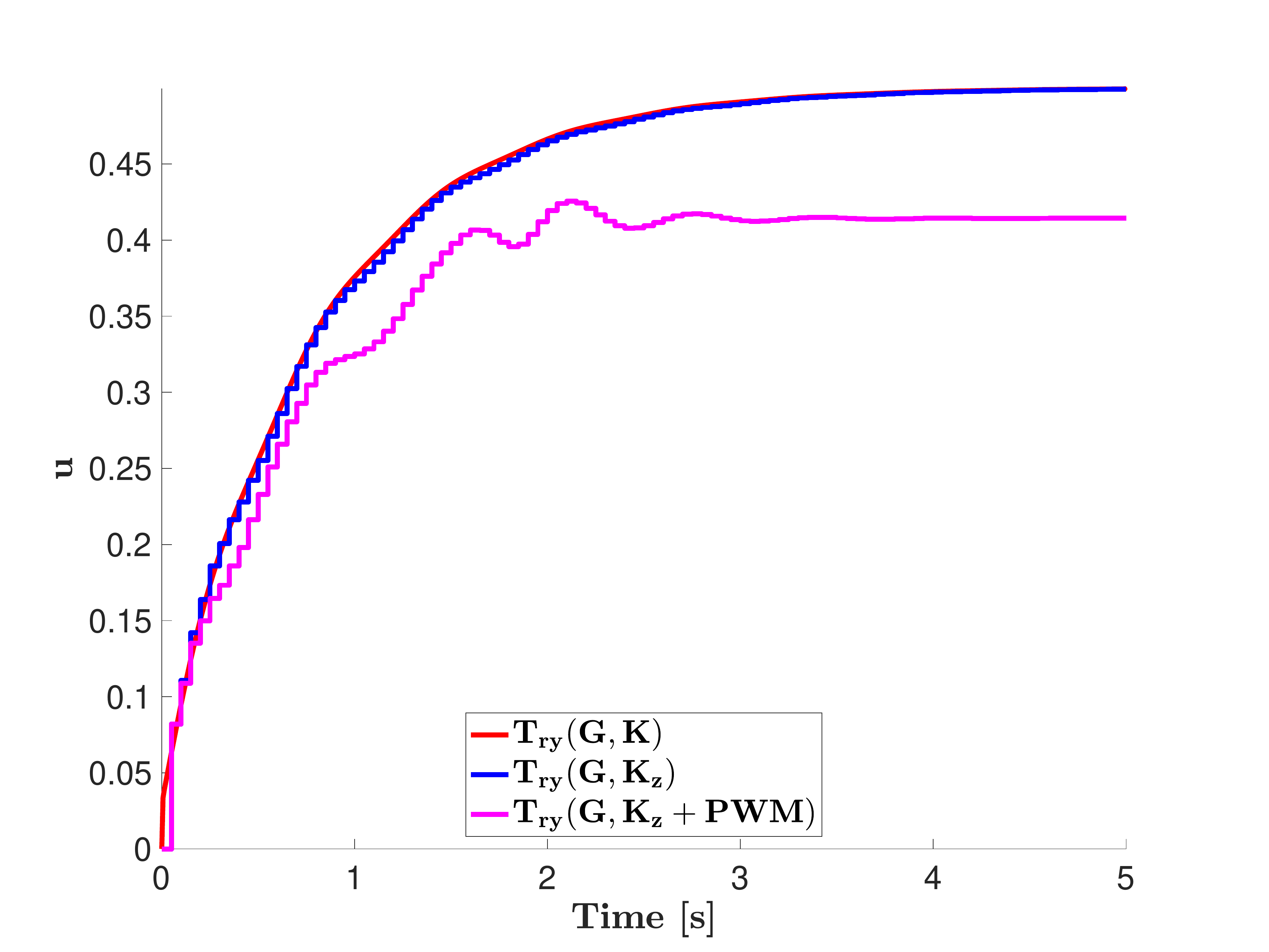}
    \includegraphics[width=.45\columnwidth]{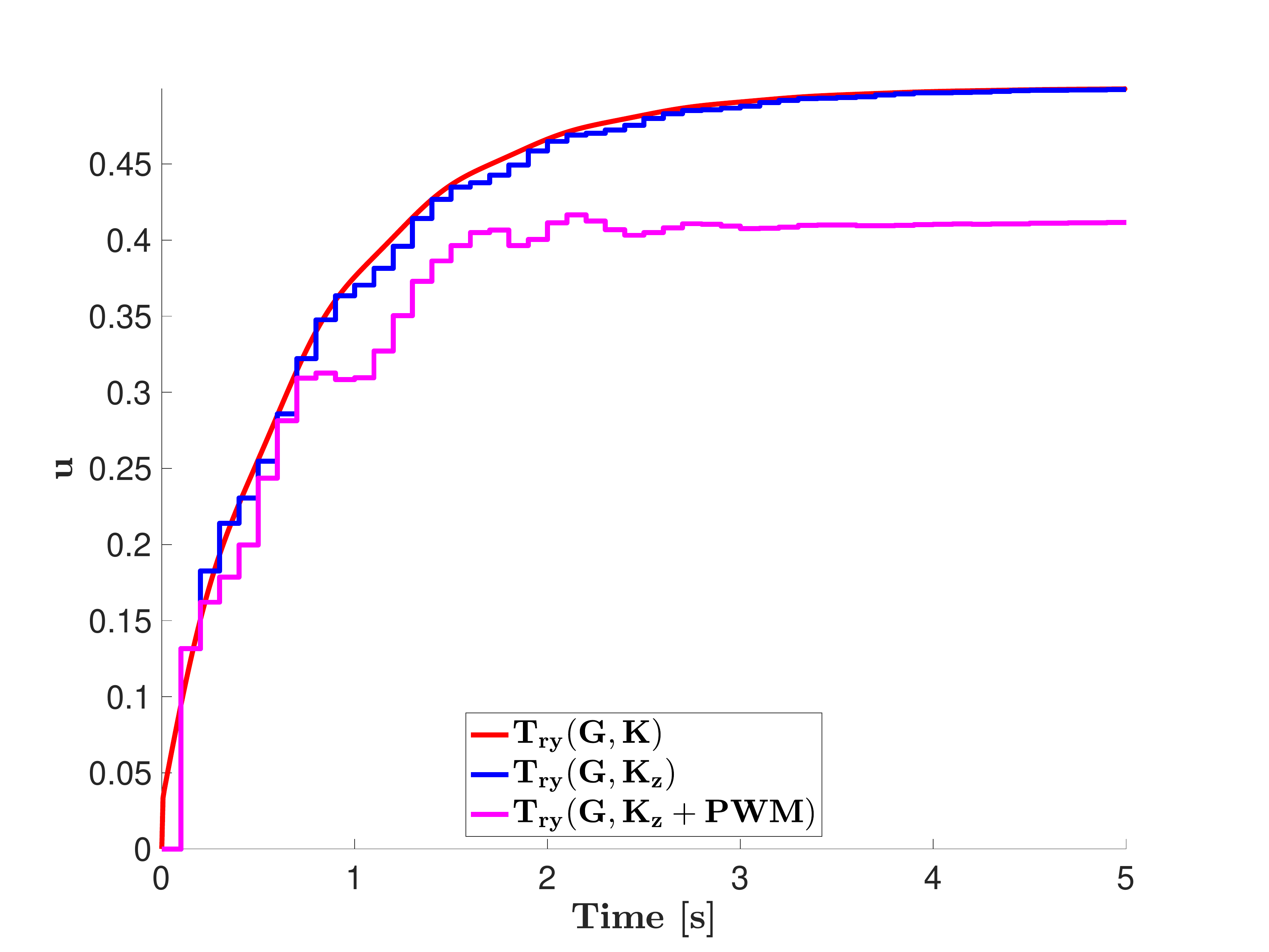}\\
    \includegraphics[width=.45\columnwidth]{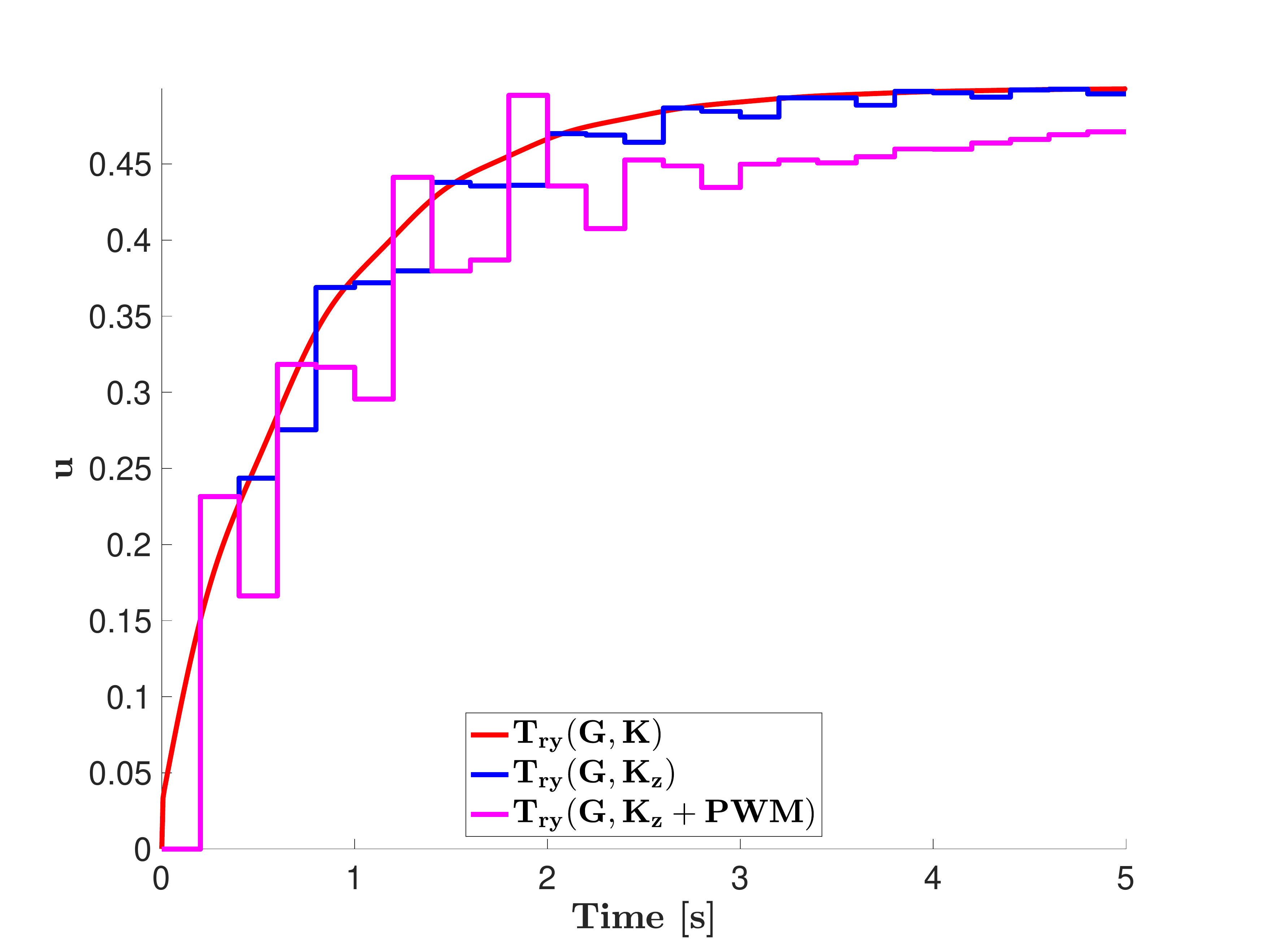}
    \includegraphics[width=.45\columnwidth]{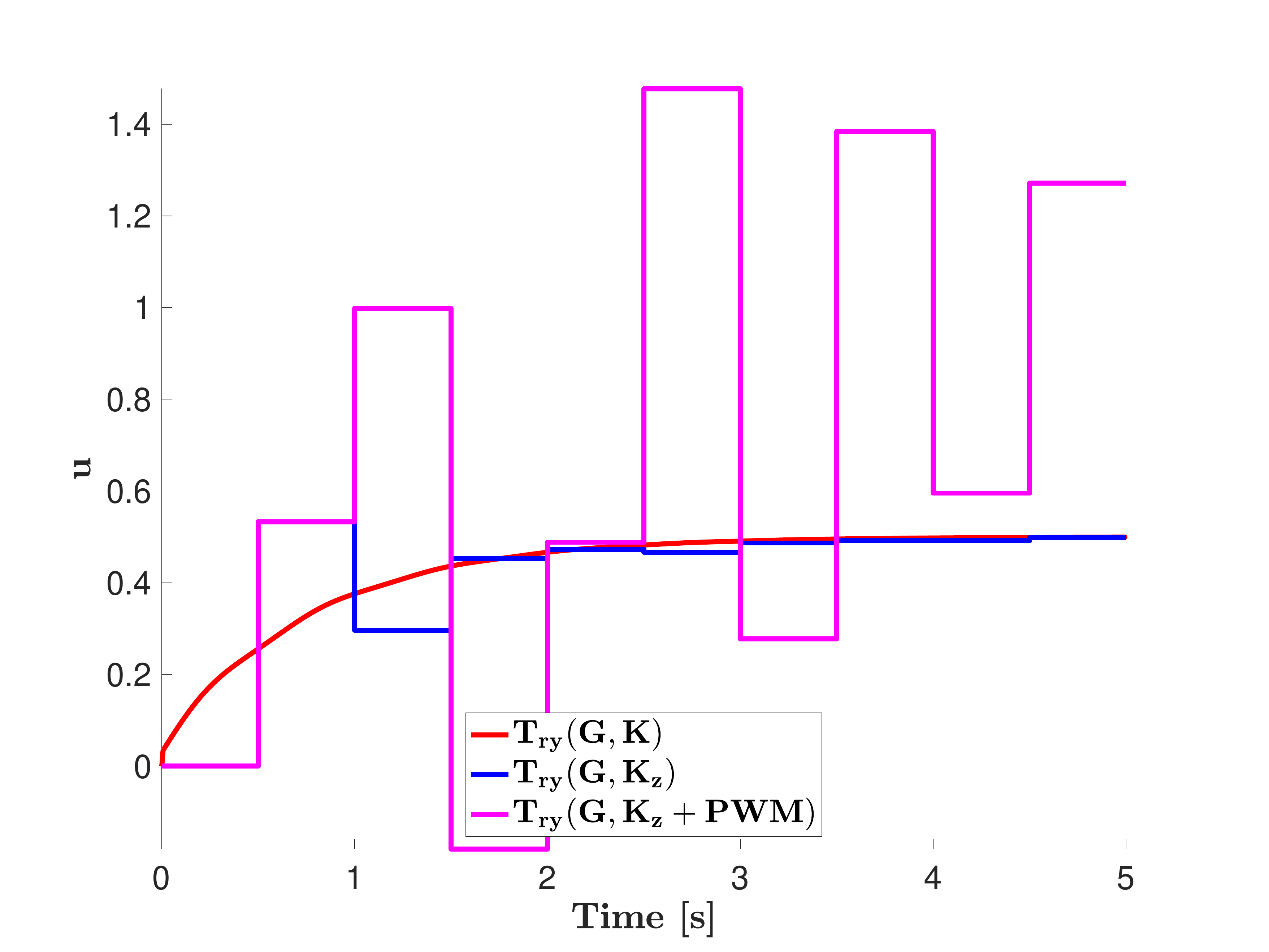}
    \caption{From top left to bottom right: $T_s=50$ms, $T_s=100$, $T_s=200$ and $T_s=500$ms.  Comparison of the controller control signal of closed-loop scheme. The continuous-time closed-loop $\Gtran$-$\mathbf K$ (solid red) and the hybrid continuous-sampled-time $\Gtran$-$\mathbf K_z$ in  (solid blue).}
    \label{fig:odeU}
\end{figure}

\begin{figure}[H]
    \centering
    \includegraphics[width=.45\columnwidth]{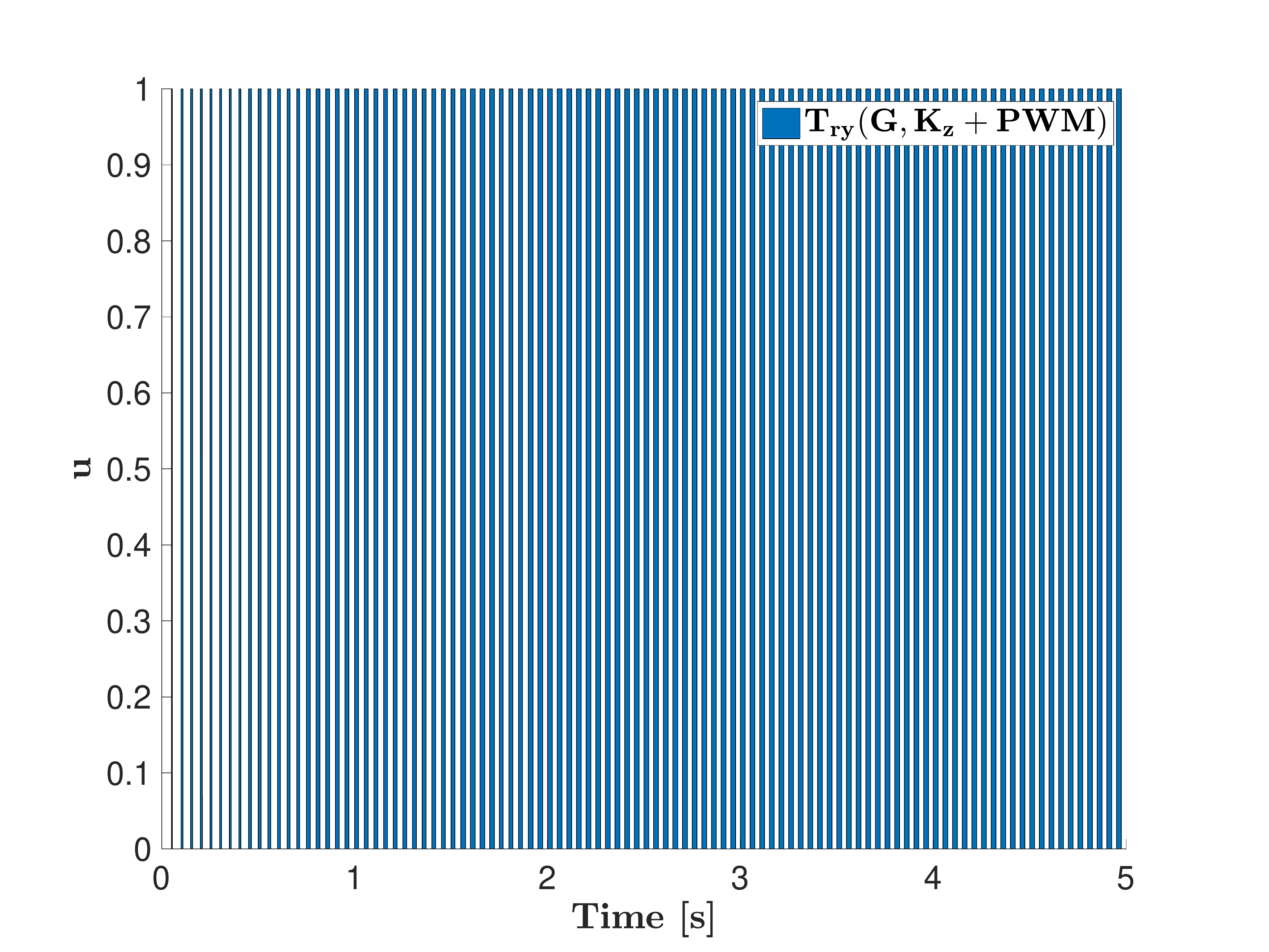}
    \includegraphics[width=.45\columnwidth]{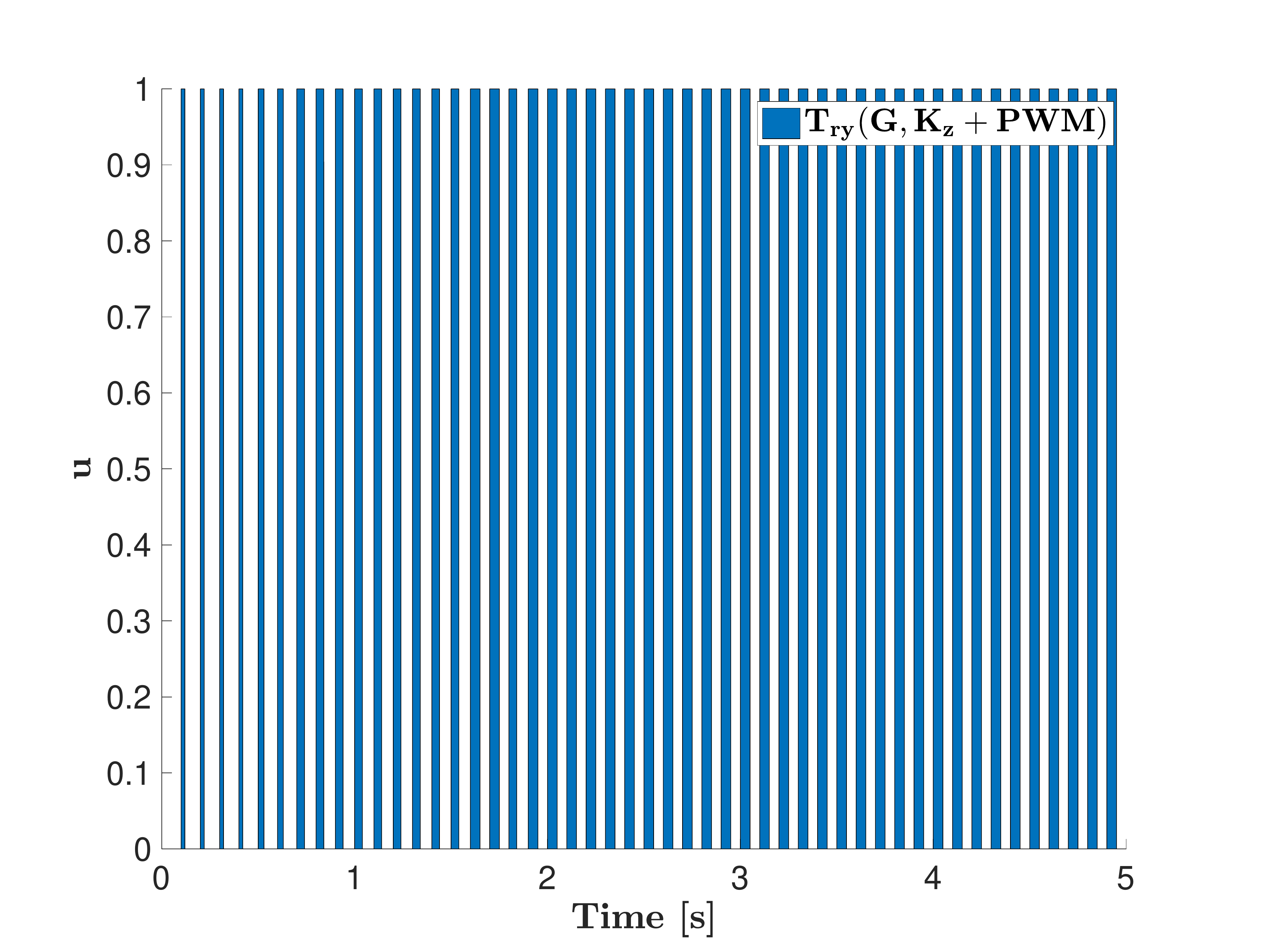} \\
    \includegraphics[width=.45\columnwidth]{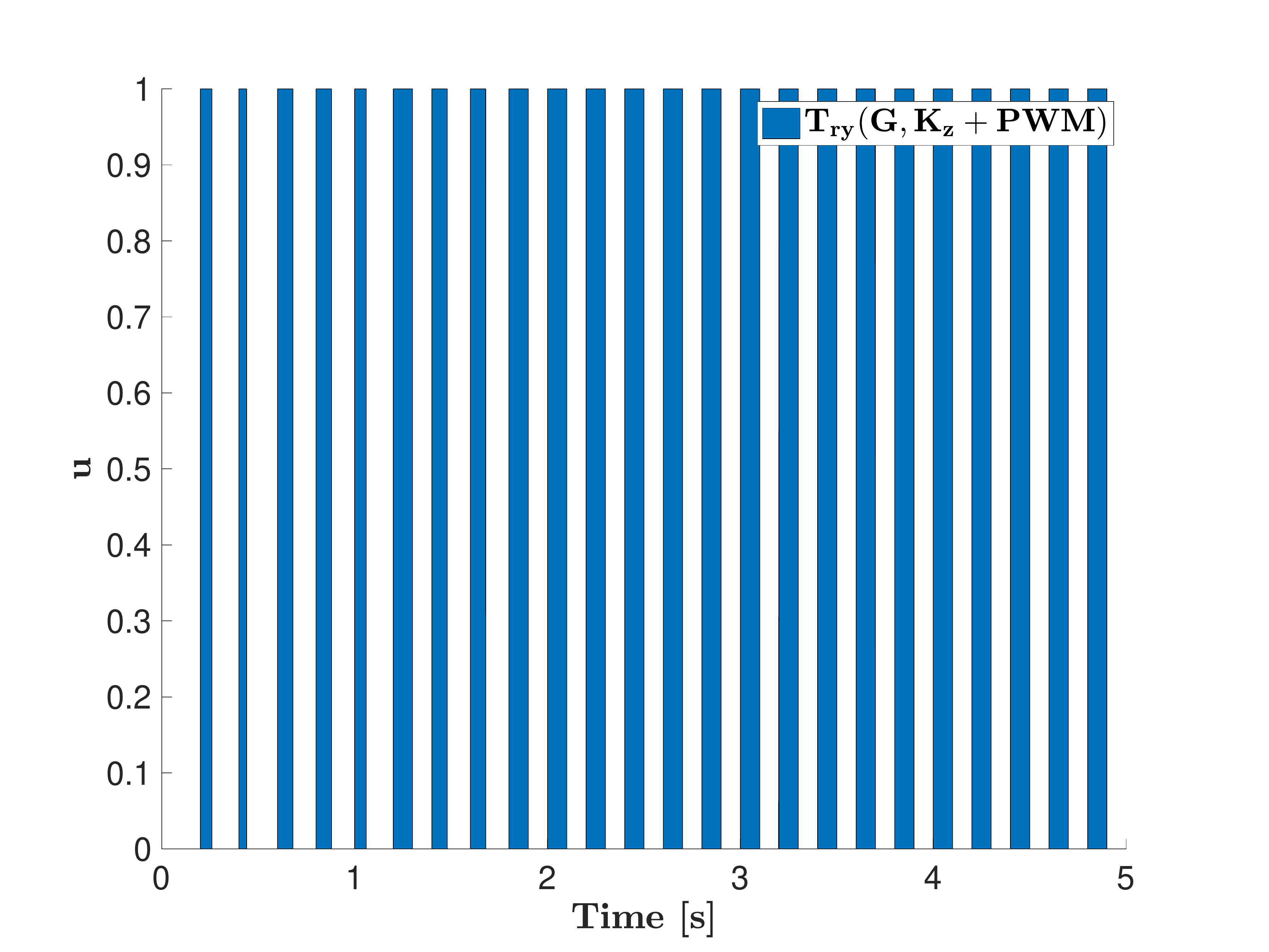}
    \includegraphics[width=.45\columnwidth]{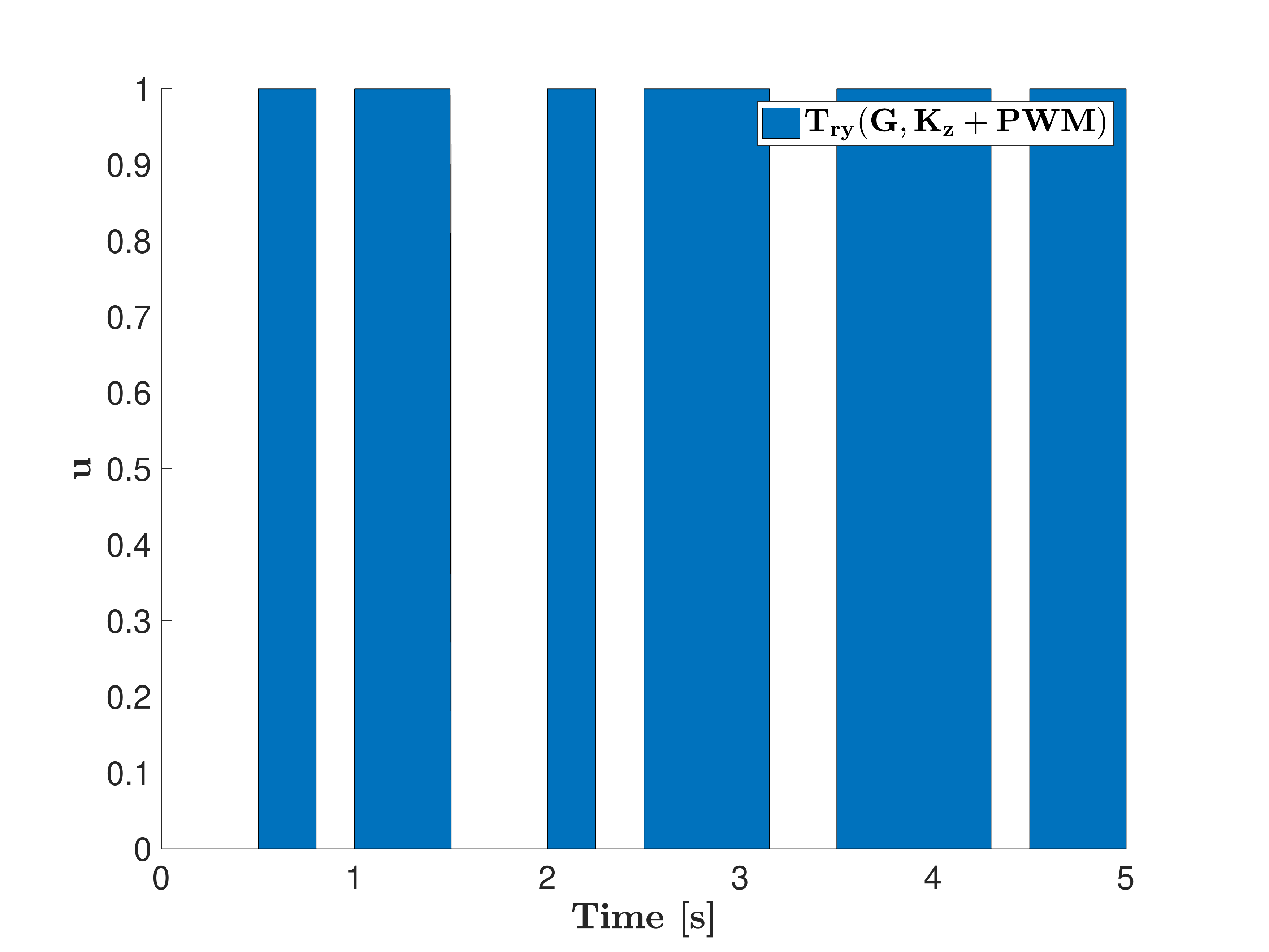}
    \caption{From top left to bottom right: $T_s=50$ms, $T_s=100$, $T_s=200$ and $T_s=500$ms.  Controller control signal of closed-loop scheme with the hybrid continuous-sampled-time and modulated $\Gtran$-$\mathbf K_z$-\textbf{PWM}. The  blue coloured areas represent the moments where the actuator is high.}
    \label{fig:odeUpwm}
\end{figure}

The above observations are completed with Figures \ref{fig:odeU} and \ref{fig:odeUpwm} illustrating the control signal sent to the system. On Figure \ref{fig:odeU}  the control signal of the discrete-time controller which sample time is being increased leads to noticeable differences from the reference continuous one. The same comment can be done on Figure \ref{fig:odeUpwm} which illustrates the pulse sent to the system. These pulses, being larger and larger due to the sampling period increased. Obviously, these observation open the field for more investigation, that will be done in the future.

%\newpage
\section{Conclusions}
\label{sec-conclusion}

In this report, we aimed at presenting in a condensed and obviously incomplete way, a standard approach for controller design and implementation. We tried to follow what authors believe is a classical control-engineer approach, starting from the system excitation, model identification and reduction, followed by a control design, and ending with some implementation issues related to a pulsed modulation-driven actuator. The report is concentrated on linear problems and nonlinear issues are not really faced here (unless the modulation part). Still, reader should keep in mind that linear dynamical systems and control methods are largely enough for many applications and mastering them is already a nice step forward.

Obviously, the report may be amended, commented and discussed according the reader knowledge, but from the past discussions authors had with multiple users, we feel that such bundle of pages can be a simple but sufficiently good starting point for many practitioners and may be useful for starting discussions.

The objective of the authors was to provide a didactic overview for unfamiliar authors by providing a step by step overview of the general ideas, accompanied with \matlab code involving the \mor Toolbox and some functions available on demand.

%%%%%%%%%%%%%%%%%%%%%%
\bibliographystyle{plain}
%\bibliography{_biblioCPV,_biblioPoussot}

\end{document}